\documentclass[9pt,twocolumn,twoside]{revtex4-1}

\usepackage{amsmath}    % need for subequations
\usepackage{amssymb}   
\usepackage{amsfonts}   
\usepackage{graphicx}   % need for figures
\usepackage{verbatim}   % useful for program listings
\usepackage{color}      % use if color is used in text
\usepackage{subfigure}  % use for side-by-side figures
\usepackage{hyperref}   % use for hypertext links, including those to external documents and URLs
\usepackage{graphics}
\usepackage[version=4]{mhchem}

\begin{document}

\preprint{AIP/123-QED}

\title{Continuous attractor-based clocks are unreliable phase estimators}

\author{Weerapat Pittayakanchit}
\thanks{WP and ZL contributed equally to this work.}
\affiliation{Department of Physics and the James Franck Institute, University of Chicago, Chicago, IL.}
\author{Zhiyue Lu}
\thanks{WP and ZL contributed equally to this work.}
\affiliation{Department of Chemistry and the James Franck Institute, University of Chicago, Chicago, IL.}
 \author{Justin Chew}
\affiliation{Medical Scientist Training Program, Pritzker School of Medicine, University of Chicago,Chicago, IL.}
\affiliation{Department of Molecular Genetics and Cell Biology, University of Chicago, Chicago, IL.}
\author{Michael J. Rust}
\affiliation{Department of Molecular Genetics and Cell Biology, University of Chicago, Chicago, IL. }
\affiliation{Department of Physics, University of Chicago, Chicago, IL. }
\author{Arvind Murugan}
 \email{To whom correspondence should be addressed. E-mail: amurugan@uchicago.edu}
\affiliation{Department of Physics and the James Franck Institute, University of Chicago, Chicago, IL. }

\date{\today}
%\keywords{estimators $|$  circadian clocks$|$ cyanobacteria $|$ internal noise $|$ attractors} 

\begin{abstract}
Statistical estimation theory determines the optimal way of estimating parameters of a fluctuating noisy signal. However, if the estimation is performed on unreliable hardware, a sub-optimal estimation procedure can outperform the previously optimal procedure. Here, we compare classes of circadian clocks by viewing them as phase estimators for the periodic day-night light signal. We find that continuous attractor-based free running clocks, such as those found in the cyanobacterium \textit{Synechococcus elongatus} and humans, are nearly optimal phase estimators since their flat attractor directions efficiently project out light intensity fluctuations due to weather patterns (`external noise'). However, such flat directions also make these continuous limit cycle attractors highly vulnerable to diffusive 'internal noise'. Given such unreliable biochemical hardware, we find that point attractor-based damped clocks, such as those found in a smaller cyanobacterium with low protein copy number, \textit{Prochlorococcus marinus}, outperform continuous attractor-based clocks. 
By interpolating between the two types of clocks found in these organisms, we demonstrate a family of biochemical phase estimation strategies that are best suited to different relative strengths of external and internal noise.
\end{abstract}

\maketitle

Extracting information from a noisy external signal is fundamental to the survival of organisms in dynamic environments \cite{Bowsher2014-xx}. 

From yeast anticipating the length of starvation \cite{Mitchell2015-oa} and bacteria estimating the availability of sugars\cite{Sourjik2012-fc,Tu2008-dm}, to dictyostelium counting the number of cAMP pulses \cite{Cai2014-ca}, organisms must often filter noisy irregular aspects of the environment while inferring parameters about a regular aspect in order to be well-adapted \cite{Siggia2013-la,Mora2010-tu, Endres2009-ft}.  

A striking example of regularity in environmental stimuli is the daily day-night cycle of light on earth; organisms from all kingdoms of life use circadian clocks to estimate the phase of these periodic signals of fixed frequency in order to anticipate and prepare for future changes in light \cite{Winfree2001-pr}. Phase inference on such an environmental signal is a challenge because unrelated aspects of the signal, such as large amplitude fluctuations due to weather patterns are uninformative of phase but the entrainment mechanisms looking for dawn-dusk transitions might conflate such fluctuations with true variation in phase. Poor phase entrainment is associated with a host of fitness costs in plants, rodents and humans\cite{Woelfle2004-bc}.

Algorithms to infer the phase of a periodic but noisy signals have been studied extensively in statistics \cite{Quinn2001-um,Liao2011-ea}; for example, the Bayesian theory of estimators develops optimal estimation procedures such as Maximum Likelihood Estimators (MLE) that account for prior expectations about the external signal. 
% % %

% % % 
However, in practice, the MLE may be computationally too slow or consume too much memory or other computational resources \cite{Liao2011-ea,Lovell1992-sf,Quinn2001-um}. 
Hence the engineering literature has considered `sub-optimal' alternatives for phase estimation, such as the Kay\cite{Kay1989-ua} and Tretter\cite{Tretter1985-bw} estimators, that reduce the computational complexity of the operation. 
Such sub-optimal estimators can outperform the theoretically optimal estimator when subject to time, energy or other resource constraints.
% % % 

\begin{figure*}[!htbp]
		\centering
		\includegraphics[width=0.8\linewidth]{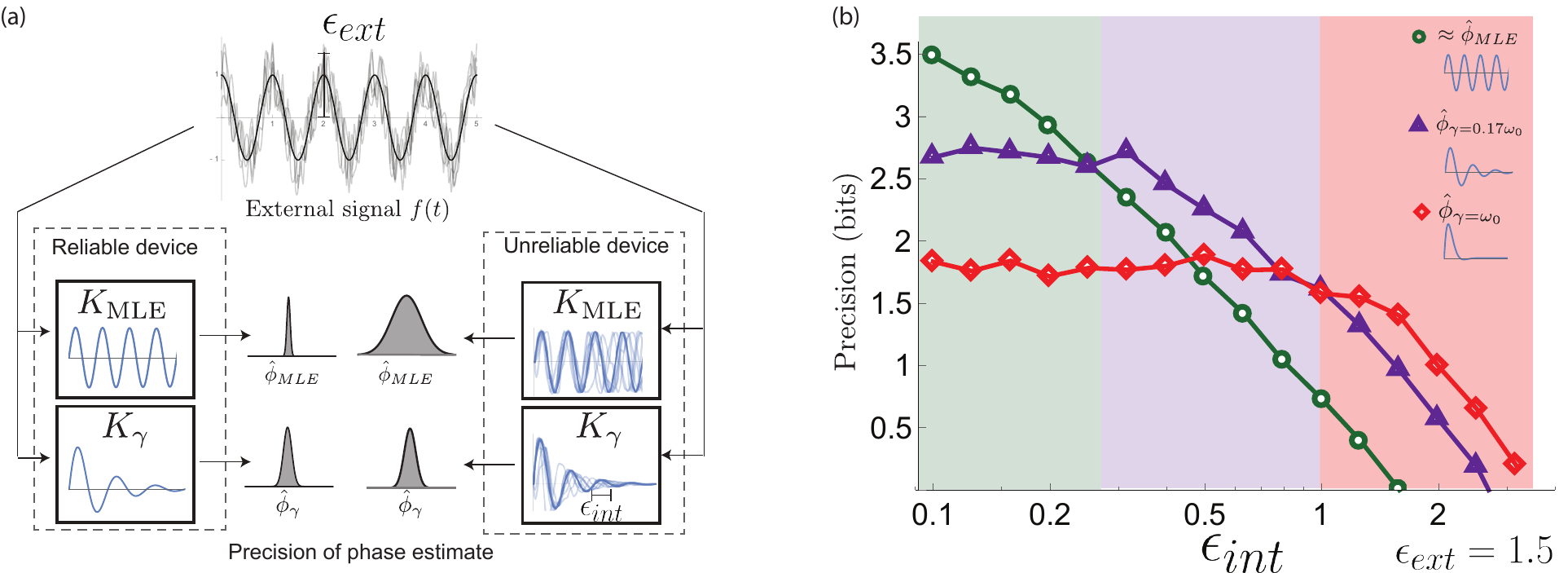}
		\caption{A sub-optimal phase estimation procedure may outperform an optimal one when both are carried out on an unreliable device. (a) The Maximum Likelihood Estimator $\hat{\phi}_{\text{MLE}}$ for the phase of a periodic signal $f(t)$ with gaussian white noise of strength $\epsilon_{ext}$ has higher precision (i.e., lower variance) than all other phase estimation strategies $\hat\phi_\gamma$ on a reliable device. Here, $\cos \hat\phi_{\gamma}(t) \propto \int^t_{-\infty} f(\bar{t}) K_\gamma(t-\bar{t}) d\bar{t}$ with $K_\gamma(t) = e^{-\gamma t} \sin \omega t$ and $K_{\text{MLE}} =\sin \omega t$. If computed on an unreliable device where the internally-generated $K_\gamma(t)$ drifts in phase over time, $\hat\phi_{\gamma}$ can have lower variance $\sigma_{\hat\phi}^2$ than $\hat\phi_{\text{MLE}}$. (b) While all estimators degrade in precision (defined as $-\log_2 \sigma_{\hat\phi}$) past a threshold of internal unreliability $\epsilon_{int}$, $\hat{\phi}_{\text{MLE}}$ is the most fragile (\textit{green}). Conversely, the overdamped estimator (\textit{red}) has poor precision on reliable hardware but outperforms all other estimators at high values of internal noise $\epsilon_{int}$.  In general, the optimal estimator $\hat\phi_{\gamma}$ for a given mix of external $\epsilon_{ext}$ and internal $\epsilon_{int}$ has $\gamma \sim \epsilon_{int}/\epsilon_{ext}$.
\label{fig:unreliableestiamtors}}
\end{figure*}

% Point 3
Molecular biology presents a novel kind of constraint on estimators, since any estimation procedure must be carried out on intrinsically unreliable biochemical hardware.  This raises the question of which estimation procedures are compatible with biophysical constraints such as finite copy number fluctuations and limited energy and time.

% Results:
Here, we evaluate the performance of a general family of circadian clocks as phase estimators of the external day-night light cycle with weather-related amplitude fluctuations; however, these estimators are intrinsically unreliable, e.g,. due to finite copy number fluctuations. Our family interpolates between free-running limit cycle clocks, like those found in humans and \textit{S. elongatus}, a $3 \mu m$ cyanobacterium, and the damped point-attractors that describe the clock in \textit{P. marinus}, a $0.5 \mu m$ cyanobacterium with an estimated $50 \times$ smaller protein copy number than \textit{S. elongatus} \cite{Bryant2003-zw,Gutu2013-oy, Holtzendorff2008-bj, Dufresne2003-gh, Kitayama2003-na}. 

We find that continuous attractors, such as limit cycles, are a double edged sword when viewed as statistical estimators. In the absence of internal fluctuations, the off-attractor dynamics of continuous attractors can selectively project out external fluctuations and thus approach Cramer-Rao bounds on estimation. However, continuous attractors are susceptible to diffusion along the attractor itself caused by internal noise (e.g. low protein copy number \cite{Potoyan2014-so}), in which case point attractors can out-perform. Thus, we find an extension of the Laughlin principle \cite{Laughlin1981-hz} - clock dynamics must be tuned to match the expected statistics of both external and internal fluctuations.

% % % %

\section{Unreliable estimators}

We first illustrate our results in a general context. Consider the canonical problem of phase estimation for a sine wave $f(t)$ of known frequency with additive Gaussian white noise of strength $\epsilon_{ext}$, extensively studied in  statistics \cite{Liao2011-ea,Lovell1992-sf} and in engineering \cite{Fu2007-ql,Ghogho1999-ni}. 

The Maximum Likelihood Estimator (MLE) for the phase at time $t$ is \cite{Liao2011-ea} $\cos \hat\phi_{\text{MLE}}(t) \propto \int^{t}_{-\infty}  f(\bar{t}) \sin(\omega (t-\bar{t}))d\bar{t}$. To physically implement such an estimator, a device must internally generate a reference sine wave of fixed frequency $\omega$ and integrate it against the entire available history of the external signal. We contrast $\hat\phi_{\text{MLE}}$ with the family of finite-history estimators given by, 
\begin{equation}
\cos \hat\phi_{\gamma}(t) \propto \int^t_{-\infty} f(\bar{t}) K(t-\bar{t}) d\bar{t} 
\label{eqn:GammaEstimator}
\end{equation} 
where $K$ is a damped oscillatory kernel; $K(t)= \sin(\omega t) e^{-\gamma t}$.  $\hat\phi_\gamma$ only accounts for a $\sim 1/\gamma$ length of the signal $f(t)$'s history.

As shown in Fig.\ref{fig:unreliableestiamtors}, on a perfectly reliable device, $\hat\phi_{\text{MLE}}$ has lower variance $\sigma_\phi^2$ than any member of $\hat\phi_\gamma$. We then turn on internal unreliability in the form of phase diffusion (with diffusion constant $\epsilon_{int}^2/2$) in generating the oscillatory kernel $K(t)$. Fig.\ref{fig:unreliableestiamtors}b shows the precision (i.e., $-\log_2 \sigma_\phi$) of $\hat\phi_{\text{MLE}}$ and two estimators in the $\hat\phi_{\gamma}$ family as a function of $\epsilon_{int}$; $\hat\phi_{\text{MLE}}$'s precision is especially fragile to internal noise.

On the other hand, estimators $\hat\phi_{\gamma=0.17 \omega},\hat\phi_{\gamma=\omega}$ based on shorter-lived kernels, are much more robust to phase diffusion and thus outperform $\hat\phi_{\text{MLE}}$ on sufficiently unreliable hardware.

Intuitively, integrating a longer history of $f(t)$, as in $\hat\phi_{\text{MLE}}$, averages out external noise but also increases exposure to internal phase drift in $K(t)$. In fact, we show in the SI that the estimator with $\gamma_{opt} \sim \epsilon_{int}/\epsilon_{ext}$ strikes the right balance in integration time and has the highest precision in this family.

\begin{figure*}[!htbp]
		\centering
		\includegraphics[width=0.9\linewidth]{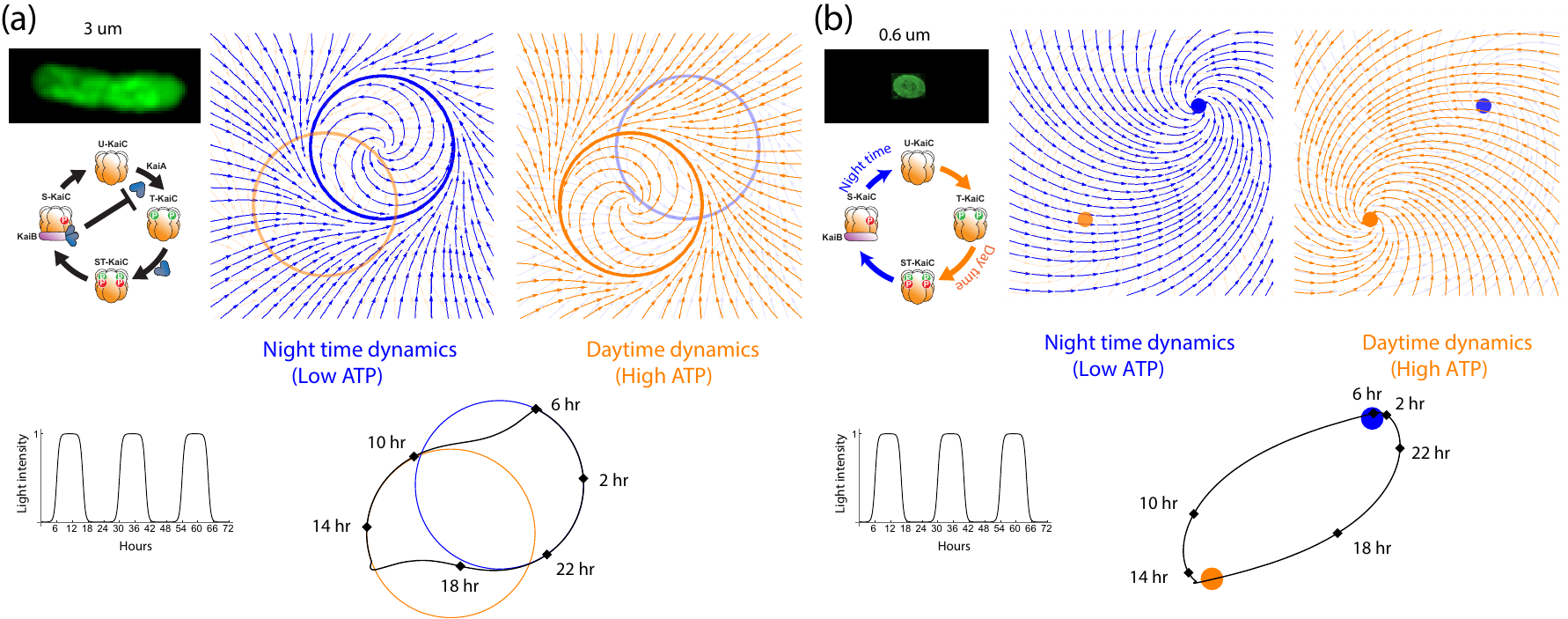}
		\caption{Circadian clocks with different dynamics are equally precise time keepers under ideal noiseless conditions. (a) Free running circadian clocks, such as the KaiABC protein clock in \textit{S. elongatus}, can be described by one limit cycle attractor during the day (orange) and a shifted limit cycle (blue) at night due to changing $ATP/ADP$ levels; the clock relaxes from one attractor to the other at dawn and dusk. (b) Damped circadian clocks, such as that in \textit{P. marinus} which lacks Kai A, are described by a day and a night point attractor with slow relaxation between the attractors over the course of the day and night. (Minimal KaiBC model consistent with \textit{P. marinus} experiments \cite{Holtzendorff2008-bj} is shown.) Under ideal circumstances - i.e., no external weather-related or any internal fluctuations - both dynamics stably entrain to the external signal, giving rise to distinct clock states that can be used a reliable internal proxy for the distinct times of the day.} \label{fig:twodynamics}
\end{figure*}

\section{Circadian clocks as estimators}

We now discuss two qualitatively distinct phase estimation strategies implemented by organisms with circadian clocks that face both external and internal fluctuations. Many organisms like humans and rodents have free running clocks that show self-sustained 24 hr rhythms even in constant dark or constant light conditions. Such clocks are phenomenologically well-described by a limit cycle attractor, a non-linear oscillator with a fixed amplitude\cite{Winfree2001-pr}. The molecular details of such limit cycle attractors are best understood for the post-translational Kai ABC protein clock in \textit{S. elongatus}; for example, the axes of the phase portrait in Fig.\ref{fig:twodynamics} could be the phosphorylation extent of the $S$ and $T$ sites on KaiC (\cite{Rust2007-op} and SI). The clock follows distinct limit cycle dynamics during the day and night\cite{Leypunskiy2017-al,Pattanayak2014-bv}, with the day cycle positioned at higher phosphorylation levels due to higher ATP levels. 

We model such free-running clocks using circular day and night limit cycles of radius $R$ in a plane. The limit cycle is defined by the dynamics $\tau_{relax} \dot{r} = r - r^3/R^2 , \dot{\theta} = \omega $ about its own center; but the center of the limit cycle itself moves along the $y=x$ diagonal in Fig.\ref{fig:twodynamics}a as $(-\rho(t) L,-\rho(t) L)$ where $\rho(t) \in [0,1]$ is the normalized light level at time $t$ and $L$ is a measure of the physiological changes between day and night (e.g., ATP/ADP ratio change in \textit{S. elongatus}). Thus, e.g. in Fig.\ref{fig:twodynamics}a, the system follows the blue dynamics at night and then after dawn it relaxes to the orange day attractor on a time scale $\tau_{relax}$. In reality, the day and night limit cycles are not circles of the same size in a plane and physiological changes might lag light levels; we later use a molecular model of the KaiABC oscillator that violates all these assumptions about shape, size and relaxation to show that our qualitative results do not rely on these assumptions. We do not include transcriptional coupling \cite{Zwicker2010-de, Paijmans2016-hs} of the clock here. Other biological oscillators described by our picture of limit cycle include NF-$\kappa$B \cite{Potoyan2014-so} driven by TNF changes \cite{Heltberg2016-ot}, and synthetic oscillators \cite{Potvin-Trottier2016-bz,Tsai2008-gg,Elowitz2000-lz}. 

Not all organisms have a free-running clock; for example, many insects \cite{Saunders2002-hj} have damped `hourglass' clocks that decay to a fixed point under constant light or constant dark conditions but show oscillatory dynamics under day-night cycling. In fact, a sister cyanobacterial species \textit{Prochlorocaucus marinus} has a KaiBC-protein based clock without the negative KaiA-feedback \cite{Dufresne2003-gh, Holtzendorff2008-bj}. Consequently, in constant light or constant dark conditions, the clock's state decays to a distinct day or a night state respectively \cite{Holtzendorff2008-bj}. Such clocks are phenomenologically well-described by a day-time and a night-time point attractor with slow relaxation dynamics between them as shown in Fig. \ref{fig:twodynamics}b, modeled as $\dot{r} = - r/\tau_{{relax}} , \dot{\theta} = \omega$ about an attractor point whose location varies with current light levels as $(-\rho(t) L,\rho(t)L)$. Here we assume $2 \tau_{relax} \sim 24$ hrs as in \textit{P. marinus} \cite{Holtzendorff2008-bj}; if relaxation were faster and completed before the day is over, the clock cannot resolve all times of the day.

With cloudless day-night cycling, both kinds of clocks entrain into a stable trajectory as shown in the lower panels of Fig.\ref{fig:twodynamics}a and b, switching dynamics between the two limit cycles or point attractors at dawn and dusk.  In what follows, we will also consider a family of limit cycle clocks of varying $R/L$ to interpolate between large-$R/L$ limit cycles and point attractors. The Hopf bifurcation is the simplest way to parametrize such an interpolation \cite{Winfree2001-pr,Murayama2017-vj}. However, the relaxation time $\tau_{relax}$ changes dramatically near a Hopf bifurcation, distracting from the effects of noise that we wish to study. Hence we hold $\tau_{relax}$ fixed in the interpolation but stop at a non-zero $R/L$ to avoid singularities (see SI).

\begin{figure*}[!htbp]
		\centering
		\includegraphics[width=\linewidth]{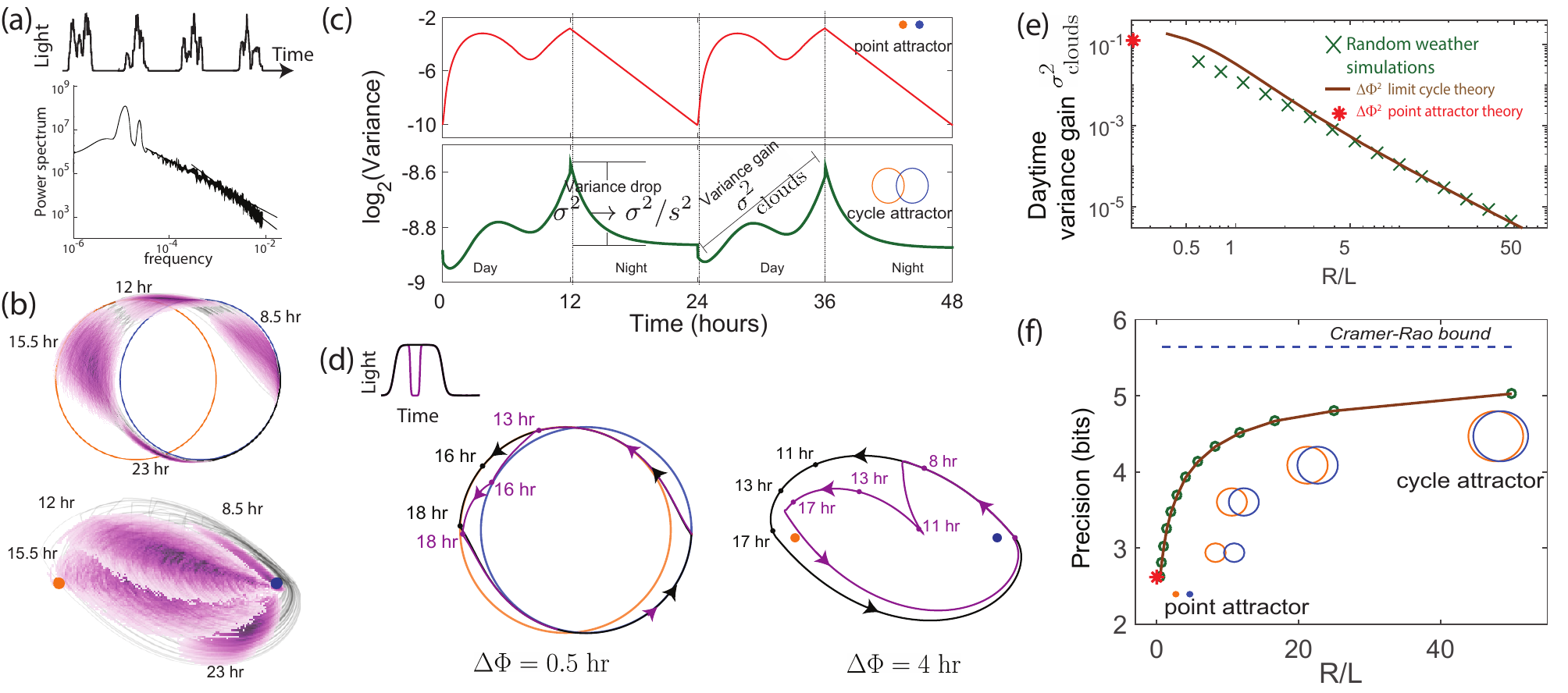}
		\caption{External weather-related light fluctuations are filtered out by limit cycle attractors but not by point attractors. (a) Light intensity levels fluctuate on a range of time scales due to weather (power spectrum reproduced from \cite{Gu2001-re}). (b) A population of limit cycle clocks of identical fixed geometry, subject to different realizations of weather conditions, show non-overlapping distributions (purple blobs) at different times of the day. Point attractor clocks form larger and more overlapping distributions. (c) In both cases, the population variance grows $\sigma^2 \to \sigma^2 + \sigma^2_{\text{clouds}}$ during the day (6 - 18 hr) and shrinks $\sigma^2 \to \sigma^2/s^2$ at dawn (6 hr) and at dusk (18 hr). (d) A single representative dark pulse, of $\sim 2.4 $ duration in this figure, causes only a $\Delta \Phi \sim 30$ min phase lag in limit cycles but $\Delta \Phi \sim 4$ hr for point attractors. The differing impact is because the clock trajectory's deviation (purple) is fundamentally bounded by the separation of the line-like attractors, in contrast to the free-fall towards the blue night-time attractor for point attractors. (e) The geometrically computed $ \Delta \Phi^2$ phase shift for a dark pulse of any fixed duration and time of occurrence (see SI) drops rapidly as $(R/L)^{-2}$ for large-$R/L$ limit cycles; this theoretical prediction agrees well with the population variance gain over the day $\sigma^2_{\text{clouds}}$ seen in panel (c). (f) Consequently, large-$R/L$ limit cycles can tell time with higher precision, asymptotically reaching the Cramer-Rao bound on optimal estimators.} \label{fig:inputnoise}
\end{figure*}

\begin{figure*}[!htbp]
		\centering
		\includegraphics[width=\linewidth]{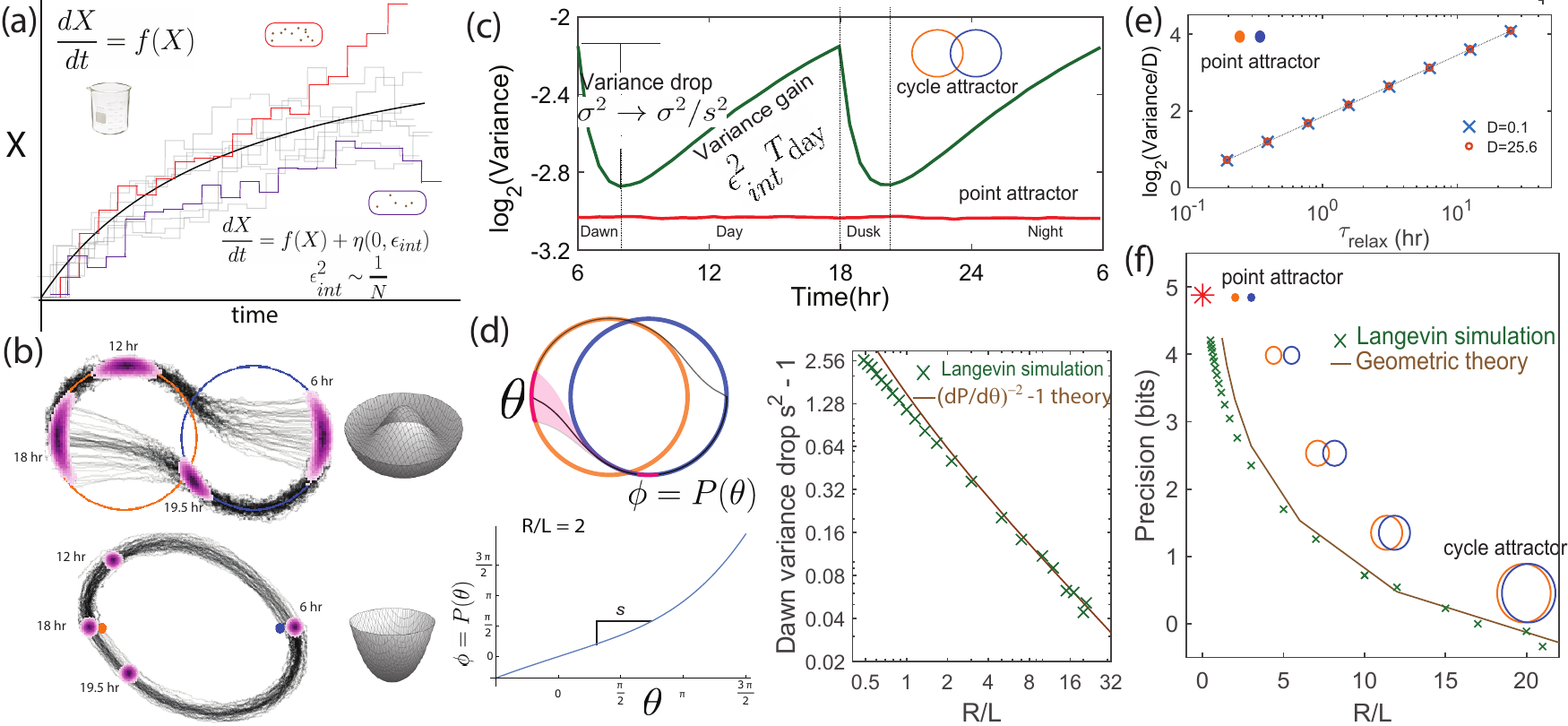}
		\caption{Internal fluctuations severely affect continuous attractors but not point attractors. (a) We model fluctuations due to finite copy number $N$ as Langevin noise $\eta(0,\epsilon_{int})$, resulting in a diffusion constant $\epsilon_{int}^2 \sim 1/N$ for the clock state. (b,c) The flat direction of limit cycles cannot contain diffusion, leading to large increases $\epsilon_{int}^2 T_{day}$ in population variance of clock state during each day (and night). In contrast, point attractor dynamics have constant curvature at all times, leading to a constant population variance over time. (d) The variance drops $\sigma^2 \rightarrow \sigma^2/s^2$ at dawn and dusk for limit cycles during the off-attractor dynamics between the day and night cycles. As with external noise, the variance drop is predicted by the slope $dP(\theta)/d\theta$ of the circle map between the cycles. This dawn/dusk drop goes to zero for large $R/L$ limit cycles but variance still increases during the day and night. (e) The variance for point attractors is $D \tau_{relax}$, a constant determined by the curvature $\tau_{relax}^{-1}$ of the harmonic potential. (f) Thus, with only internal noise present, the precision of limit cycle clocks increases with increasing separation $L/R$, asymptotically approaching the performance of point attractors. } \label{fig:internalnoise}
\end{figure*}

\section{External noise - weather patterns}

We begin with the performance of different clocks in the presence of external intensity fluctuations due to weather patterns. Weather patterns cause large fluctuations in the intensity of light over a wide range of time-scales as shown in Fig.\ref{fig:inputnoise}a. 

We model such fluctuations during the day as random dark pulses that cause a temporary shift back to the night cycle dynamics. In what follows, we quantify time-telling precision of clocks by first subjecting an \textit{in silico} population of bacteria to different realizations of such noisy weather patterns. We compute the resulting distribution $p(\vec{c}|t)$ in clock state $\vec{c}$ at a given time of day $t$. The variance of $p(\vec{c}|t)$ fundamentally limits the precision with which the cell can infer the current time $t$ from the clock state. Finally, we average variance over the day-night cycle to find mutual information between clock state and time \cite{Tostevin2009-cm}. See SI. Alternative related measures include the ability to anticipate sunset or sunrise.

When subject to weather fluctuations, we see in Fig.\ref{fig:inputnoise}b that the population variance of clock states for limit cycles is fundamentally limited by the spacing between the day and night limit cycles. Point attractors develop much larger and overlapping population distributions at different time points.  

We can geometrically understand the daytime variance increase $\sigma^2_{clouds}$, seen in Fig.\ref{fig:inputnoise}c, in terms of the phase lag $\Delta \Phi$ due to a single, say $2.4$ hr dark pulse \cite{Winfree2001-pr} administered during the day.  Fig.\ref{fig:inputnoise}d shows that the deviation in trajectory for limit cycle clocks (purple) is fundamentally limited by the presence of the two continuous attractors.  In contrast, for the point attractor, a dark pulse sets the system in free fall towards the night point attractor, with no limit cycle to arrest such a fall. Consequently, the geometrically computed phase shift $\Delta \Phi$ due to the particular dark pulse shown in Fig.\ref{fig:inputnoise}d is much smaller for limit cycles ($\Delta \Phi \sim 0.5 $ hr for the $R,L$ geometry shown) than for point attractors ($\Delta \Phi \sim 4 $ hr) (see SI).

In fact, this contrast in $\Delta \Phi$ between limit cycles and point attractors holds for dark pulses of any duration and time of occurrence (see SI). Finally, the contrast is even greater at large $R/L$ as shown in see Fig.\ref{fig:inputnoise}e;  the dark pulse phase shift $(\Delta \Phi)^2 \sim (L/R)^2$ falls rapidly with limit cycle size. This trend agrees with the variance gain $\sigma^2_{clouds}$ seen in simulations that average over random weather conditions. Hence, large-$R/L$ limit cycles are much less affected by external fluctuations than point attractors.

To complete the analysis, note that in Fig.\ref{fig:inputnoise}c, the population variance increases additively during the day and falls multiplicatively at dusk (and dawn), i.e., $\sigma^2 \xrightarrow{day} \sigma^2 + \sigma^2_{clouds} \xrightarrow{dusk} (\sigma^2 + \sigma^2_{clouds})/s^2 \xrightarrow{night} \ldots$. Solving for steady state, we find  
\begin{equation}
\sigma^{2,ext}_{limit\; cycle} \sim \Delta \Phi^2/(s^2 - 1).
\label{eqn:extlimitcycle}
\end{equation}
where we have equated $\sigma^2_{clouds}$ to $\Delta \Phi^2$ for a typical dark pulse. We must now compute the variance drop $\sigma^2 \to \sigma^2/s^2$ seen at dusk (and dawn). As shown in the SI for external noise (and in Fig.\ref{fig:internalnoise}b for internal noise), this dawn/dusk entropy drop can be geometrically explained by the slope of the circle map relating the two cycles \cite{Leypunskiy2017-al}; we find that $s^2 -1 \sim L/R$ for large-$R/L$ limit cycles. Plugging this and $\Delta \Phi^2 \sim (L/R)^2$ into Eq.\ref{eqn:extlimitcycle}, we see that $\sigma^2 \to L/R \to 0$ for large cycles. 

Fig.\ref{fig:inputnoise}f shows that the precision (i.e., mutual information between clock state and time) computed from random weather simulations agrees with this theory; clock precision drops as we interpolate from limit cycles to point attractors. 

\section{Internal noise - finite copy number}

In addition to external fluctuations, circadian clocks must also deal with the intrinsically noisy nature of biochemical reactions\cite{Swain2002-tj}. In particular, based on their relative sizes\cite{Dufresne2003-gh, Holtzendorff2008-bj,Bryant2003-zw}, \textit{P. marinus} is thought to have far fewer copies of the Kai clock proteins (e.g., $\sim 500$ of KaiC )than \textit{S. elongatus} ($\sim O(10000)$ copies of KaiC \cite{Gutu2013-oy, Kitayama2003-na}). Such finite numbers of molecules is known to create significant stochasticity in oscillators, even in the absence of an external signal \cite{Potoyan2016-li}.

Finite copy number effects on cellular function have been extensively studied and modeled \cite{Ziv2007-ga, Mugler2010-bh,Qian2011-yc}, e.g., using Gillespie simulations. Here we follow \cite{Potoyan2014-so,Gillespie2007-yr} and add Langevin noise to all dynamical variables of the system of strength $\epsilon_{int} \sim 1/\sqrt{N}$, where $N$ is the overall copy number, with the ratios of different species assumed fixed (see SI).  
In the Langevin approach, the clock state still has dynamics implied by the phase portrait in Fig.\ref{fig:twodynamics} but also diffuses with a diffusion constant $\epsilon_{int}^2 \sim 1/N$. We later check our results against full Gillespie simulations of an explicit Kai ABC model. 

We simulated a population of clocks in externally noiseless day-night light cycles but with internal Langevin noise. We see in Fig.\ref{fig:internalnoise}b that limit cycle populations have significantly higher variance of clock state due to internal noise than point attractors, in contrast to Fig.\ref{fig:inputnoise}b with external noise alone. 

We can understand the weakness of limit cycle attractor relative to the point attractor in terms of diffusion along flat and curved directions in the phase plane. The flat direction along the limit cycle attractor cannot contain diffusion caused by the Langevin noise and hence the population variance along the limit cycle increases linearly with time during the day, changing by $\sigma^2 \to \sigma^2 + \epsilon_{int}^2 T_{day}$ during a day of length $T_{day}$ (and similarly at night), as shown in Fig.\ref{fig:internalnoise}c. 

Dawn and dusk times do reduce the variance $\sigma^2 \to \sigma^2/s^2$ as the trajectories originating on, say, the day cycle converge on the night cycle (see Fig. \ref{fig:internalnoise}d and \cite{Leypunskiy2017-al}). In fact, we can compute this variance drop $s^2$ entirely through geometric considerations. We define the circle map $\phi = P(\theta)$ as relating originating points $\theta$ near dusk on the day cycle to final points on the night cycle $\phi$ after relaxation (experimentally characterized in \cite{Leypunskiy2017-al}). Then $s^{-1} = dP(\theta)/d\theta$. Fig.\ref{fig:internalnoise}d shows that this slope $s^{-1} = dP(\theta)/d\theta$, geometrically computed in the SI, agrees with the dawn/dusk variance drop in Langevin simulations and scales as $s^2 - 1 \sim L/R$ for large $R/L$.

Thus, the population variance changes as $\sigma^2 \xrightarrow{Day} \sigma^2 + \epsilon_{int}^2 T_{day} \xrightarrow{Dusk} (\sigma^2 + \epsilon_{int}^2 T_{day} )/s^2 \xrightarrow{Night} \ldots $ where the night adds another $ + \epsilon_{int}^2 T_{night}$ and so on. Assuming $T = T_{day}=  T_{night}$ and solving for steady-state average variance,
\begin{equation}
\sigma^{2,int}_{cycle} \sim \epsilon_{int}^2 T \frac{s^2+1}{s^2 -1} 
\label{eqn:IntCycleSpread}
\end{equation}
Consequently, as the cycles become large (large $R/L$), the dawn/dusk variance drop vanishes as $s^2 -1 \sim L/R \to 0$
while diffusion along the flat direction still adds $+\epsilon_{int}^2 T$ to the variance during each day and each night; hence large-$R/L$ limit cycles have large $\sigma^{2,int}_{cycle}$ and thus low precision.

In contrast, for the point attractor, the population variance stays constant during the day-night cycle. The size of this variance is analytically shown in the SI to be, 
\begin{equation}
\sigma^{2,int}_{point} \sim \epsilon_{int}^2 \tau_{relax}
\label{eqn:IntPointSpread}
\end{equation}
which matches Langevin simulations as shown in Fig.\ref{fig:internalnoise}e. Since  $\tau_{relax} \sim T_{day}$ to have distinct clock states throughout the day (see Fig.\ref{fig:twodynamics}c)), we find $\sigma^{2,int}_{cycle} \geq \sigma^{2,int}_{point}$.

To summarize, in both cases, population variance is reduced by the geometric `curvature' of the dynamics which is set by how much nearby trajectories converge. Point attractor trajectories experience a constant curvature of $1/\tau_{relax}$, giving Eqn.\ref{eqn:IntPointSpread}. In contrast, limit cycle clocks have long periods of zero curvature along the limit cycle (day and night); such dephasing in constant conditions has been studied in circadian clocks \cite{Mihalcescu2004-ov, Barkai2000-nr, Gonze2002-rx,Cao2015-vz}, in NF-$\kappa$B \cite{Potoyan2016-li} and computed in similar fashion as in Eqn.\ref{eqn:IntCycleSpread} for phase oscillators \cite{Monti2017-dm}. Here, such variance increases are balanced only by  short periods of `curved' off-attractor dynamics at dawn and dusk, when clock must relax to the new day or night attractor (Fig.\ref{fig:twodynamics}a). Hence limit cycles under-perform point attractors if only internal noise is present.

\begin{figure*}[!htbp]
		\centering
		\includegraphics[width=1\linewidth]{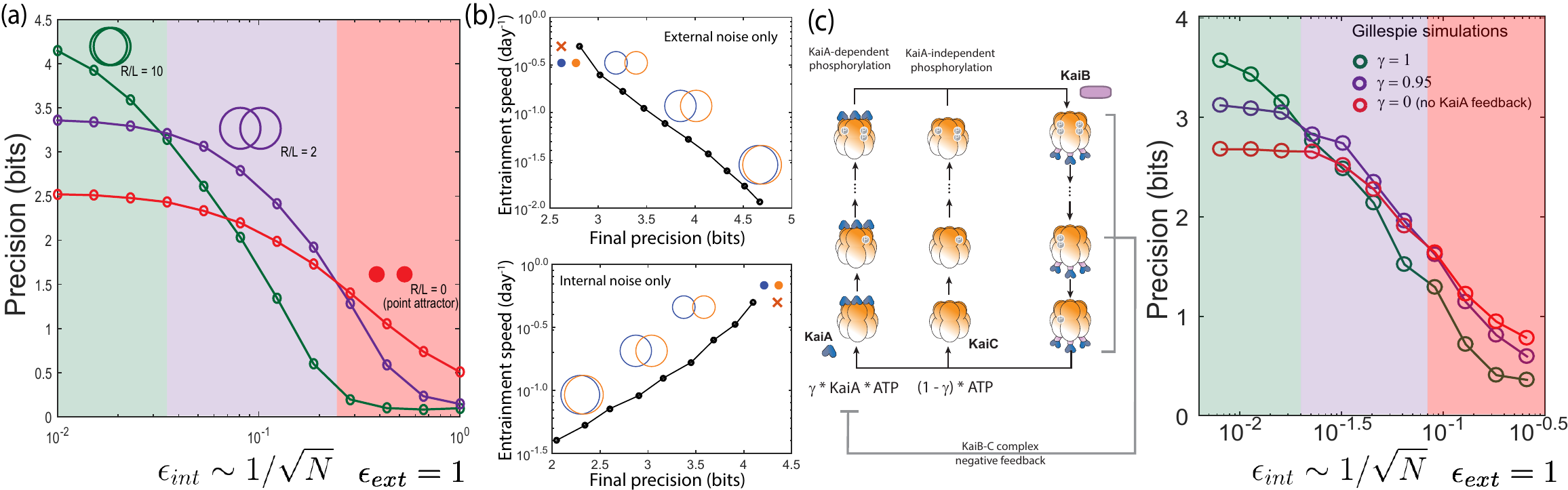}
          \caption{Large-$R/L$ limit cycle attractors outperform all other clocks in the absence of internal noise but are least robust to internal noise. (a) Point attractors and smaller $R/L$ limit cycles show low precision (i.e., low mutual information) but do not degrade as much as large-$R/L$ limit cycles with internal noise. (b) Speed-precision trade-off: With external noise alone, the most precise clocks (i.e., large $R/L$ limit cycles) are the slowest to entrain, i.e, slow to transform a population with uniform phase distribution to the steady state distribution. However, with internal noise alone, there is no trade-off between speed and precision; faster entraining clocks (i.e., point attractors) are more accurate since slow clocks are exposed to more internal noise. (c) Explicit Gillespie simulations of an explicit KaiABC clock model reproduces trends in (a). In this hybrid model, the KaiA feedback strength $\gamma$ allows interpolation between limit cycles and point attractors clocks. See SI for details. Clocks with a given $\gamma$ have highest precision in a specific range of internal to external noise strength $\epsilon_{int}/\epsilon_{ext}$.} \label{fig:intextnoise}
\end{figure*}

\section{Combination of external and internal noise}
We now subject the clock systems to both internal and external noise at the same time. We find results (see Fig.\ref{fig:intextnoise}a) that parallel those for mathematical estimators in Fig.\ref{fig:unreliableestiamtors}b. Large-$R/L$ limit cycles outperform other clocks in filtering out external noise when internal noise is low but their precision degrades more rapidly than other clocks as internal noise $\epsilon_{int}^2 \sim 1/N$ is increased. Point attractors have poor precision with only external noise but do not significantly degrade with internal noise and outperform all other clocks at high internal noise. At comparable strengths of internal and external noise, limit cycles with an intermediate value of $R/L$ are most precise.

The calculations and simulations so far assume idealized limit cycles; e.g., we assume the simplest form of circular limit cycles that exist near a Hopf bifurcation and assume the same diffusion constant $\epsilon_{int}^2$ around the limit cycle. Real biochemical oscillators such as circadian clocks \cite{Leypunskiy2017-al}, NF-$\kappa$B \cite{Potoyan2014-so}, or synthetic circuits \cite{Tsai2008-gg,Elowitz2000-lz} can violate such assumptions. To test if our results survive the specifics of biological clocks, we performed Gillespie simulations for a explicit model of KaiABC that interpolates between the known biochemistry\cite{Rust2007-op} of \textit{S. elongatus}'s clock and the putative KaiBC clock\cite{Bryant2003-zw,Holtzendorff2008-bj} in \textit{P. marinus} (Fig.\ref{fig:intextnoise}c). The limit cycles in this model are not perfect circles of the same size, do not lie entirely in two dimensions and are affected by finite copy number in a heterogeneous way (see SI). Despite such complications, we find the general behavior of Fig.\ref{fig:intextnoise}b is reproduced by this model in Fig.\ref{fig:intextnoise}c. We dial the strength of the KaiA feedback $\gamma$, responsible for spontaneous oscillations, to interpolate between limit cycles and point attractors. As earlier, we find that different ratios of internal to external noise require different strength of the KaiA feedback for highest clock precision.

\subsection{Speed-precision trade-off} 

Thus far, we have only considered the population variance at steady state as a proxy for clock quality. An independent measure of the clock quality is the entrainment speed, i.e., the time taken to reach steady state population variance, starting from a population uniformly distributed in clock phase. In Fig.\ref{fig:intextnoise}b, we show the resulting trade-off between precision and speed for our family of estimators in the presence of only external noise and then, only internal noise. With external noise, the most precise estimators (i.e., large-$R/L$ limit cycles) take much longer to reach such a steady state. Intuitively, limit cycles retain a longer history of the external signal, allowing them to average out external noise better, much like the (slow) Maximum Likelihood Estimator (Fig.\ref{fig:unreliableestiamtors}). In contrast, point attractors have little memory of the external signal seen in earlier days since the population converges to a point every night.

Strikingly, such a trade-off between speed and accuracy is absent if only internal noise is present; the estimators most robust to internal noise (i.e., point attractors) are also the fastest estimators, much as we found for statistical estimators. Intuitively, the less time spent estimating using unreliable hardware gives less opportunity for error. As with statistical estimators, with both kinds of noise present, clocks with intermediate entraining speed will have the highest precision.  

\section{Discussion}
Parameter estimation is known to be aided by having an internal model of the expected signal since external fluctuations inconsistent with that model can then be projected out easily\cite{Sontag2003-lk}. Here, we reconceptualize circadian clocks as phase estimators for noisy input signals and note that limit cycle-based free running circadian clocks encode an internal model of the expected external day-night cycle of light. We find that the continuous attractor underlying such a clock is able to effectively project out weather-related amplitude changes that are perpendicular to the flat direction. Similar roles for the flat direction of continuous attractors have been extensively explored in neuroscience \cite{Burak2012-bu}, e.g., for head and eye motor control  \cite{Seung2000-bk} and spatial navigation \cite{Yoon2013-nl}. However, we see here that the same flat direction becomes a vulnerability with internal fluctuations since such fluctuations cannot be restricted to be perpendicular to the attractor. Thus, when the internal model is unreliable, a simpler phase estimation procedure with no internal model provides better time keeping. 

Thus our work suggests that the damped circadian oscillator, like that in \textit{P. marinus} \cite{Bryant2003-zw}, is not merely a poor cousin of the remarkable free running oscillator found in \textit{S. elongatus}. At the low protein copy numbers in \textit{P. marinus}, such damped point attractors keep time more reliably than limit cycle clocks.  In addition to \textit{P. marinus}, damped oscillators are found elsewhere in biology \cite{Winfree2001-pr,Saunders2002-hj, Vaze2016-nj}. In fact, many limit cycle oscillators shrink down to point attractors as physiological conditions are varied, such as \textit{S. elongatus}'s clock at low temperatures\cite{Murayama2017-vj}, NF-$\kappa$B at very low or high levels of TNF$\alpha$ stimulation\cite{Heltberg2016-ot} or insect clocks in response to diet and temperature changes \cite{Saunders2002-hj,Kidd2015-cr}. Our work suggests that such families of oscillators that interpolate between limit cycles and point attractors continuously trade off protection against external fluctuations for protection against internal fluctuations.

\begin{acknowledgments}
We thank Aaron Dinner, John Hopfield, Eugene Leypunskiy, Charles Matthews, Brian Moths, Thomas Witten, and members of the Rust and Murugan labs for fruitful discussions.
\end{acknowledgments}

\appendix

\section{Statistical phase estimators}
The MLE for phase at time $t = 0$ for a periodic signal $f(t)$ of known frequency $\omega$ with additive Gaussian white noise (AGWN) has been well-studied and is known to be \cite{Liao2011-ea,Quinn2001-um}
\begin{equation}
\cos \hat\phi = \lim_{T\rightarrow \infty} \frac{1}{T}\int_{-T}^{0} f(t') \sin(\omega t') dt'
\label{eqn:MLE}
\end{equation}
A quick way to see this is to note that with Gaussian noise, the likelihood function is $\log e^{-x^2} \sim x^2$ and thus Maximum Likelihood Estimation is equivalent to least-squares minimization between the signal and a reference sine wave, $\mbox{argmin}_\phi || \sin(\omega t + \phi) - f(t)||^2$. Expanding the square, only the cross term $\int f(t) \sin(\omega t)$ survives since $\sin^2  \omega t$ and $f(t)^2$ terms integrate to constants. In this way,  Eq.\ref{eqn:MLE} can be shown \cite{Liao2011-ea,Quinn2001-um} to be the Maximum Likelihood Estimator for Gaussian white noise. 

If we perform this estimation `online' (i.e., provide a running estimate as a function of time), we can write this estimator in the more familiar kernel form,
\begin{equation}
\cos \hat\phi(t) =\lim_{T\rightarrow \infty} \frac{1}{T} \int_{-T}^{t} f(t') K_{MLE}(t-t') dt'
\end{equation}
where $K_{\text{MLE}}(t) = \sin(\omega t)$. 

Inspired by the constraints of carrying out such an estimator using a physical system with finite memory, we generalize the above MLE to a family of estimators:
\begin{eqnarray}
\cos ( \hat\phi_\gamma + \phi_0) =\int_{-\infty}^{t} f(t') K_{\gamma}(t-t') dt' \\
K_\gamma(t) =  \frac{\gamma\sqrt{\gamma^2+4\omega^2}}{\omega} e^{-\gamma t} \sin(\omega t)
\end{eqnarray}
where $\phi_0 = \arcsin {\frac{\gamma}{\sqrt{\gamma^2+4\omega^2}}}$ is an offset.

\subsection{External noise}
We model external noise as an additive Gaussian process,
\begin{equation}
f(t) = \cos(\omega t) + \eta(0,\epsilon_{ext})(t)
\end{equation}
where $\langle \eta(t') \eta(t) \rangle_{ext} = \epsilon_{ext}^2 \delta(t-t')$. To estimate the variance of the estimator, we denote $r(t) = \cos \hat\phi(t)$, and compute its autocorrelation function

\begin{widetext}
\begin{eqnarray}
\langle r(t) r(0) \rangle_{ext} &=& \int_{-\infty}^t \int_{-\infty}^0 \langle f(t_1) f(t_2) K_{\gamma}(t-t_1) K_{\gamma}(t-t_2) \rangle_{ext} {\rm d}t_1{\rm d}t_2 \\
&=&  \langle r(t) \rangle_{ext} \langle r(0) \rangle_{ext} + \int \int  \langle \eta(t_1) \eta(t_2) \rangle_{ext} K_{\gamma}(t-t_1) K_{\gamma}(t-t_2).  
\end{eqnarray}
\end{widetext}
Thus $C(t)\equiv \langle r(t) r(0) \rangle_{ext} -\langle r(t) \rangle_{ext} \langle r(0) \rangle_{ext}$ can be evaluated as
\begin{widetext}
\begin{eqnarray}
C(t) & = & \int \int  \langle \eta(t_1) \eta(t_2) \rangle_{ext} K_{\gamma}(t-t_1) K_{\gamma}(0-t_2) dt_1 dt_2 \\ \nonumber
& = & \int \int  \epsilon_{ext}^2 \delta(t_1 - t_2) K_{\gamma}(t-t_1) K_{\gamma}(0-t_2) dt_1 dt_2 \\ \nonumber
& = &  \epsilon_{ext}^2 \int_{-\infty}^0   K_{\gamma}(t-t_2)\cdot K_{\gamma}(-t_2) dt_2   \\
& = &  \epsilon_{ext}^2 \frac{\gamma e^{t \gamma }( \cos(\omega t)-\gamma\omega^{-1} \sin(\omega t))(\gamma^2+4\omega^2)}{4(\gamma^2+\omega^2)}   
\end{eqnarray}
\end{widetext}
Thus, 
\begin{equation}
\sigma^2=C(0)=\epsilon_{ext}^2 \frac{\gamma(\gamma^2+4\omega^2)}{4(\gamma^2+\omega^2)}\approx\epsilon_{ext}^2\gamma.
\end{equation}

Hence we conclude that for small $\gamma$, $$\sigma^2 \approx \epsilon_{ext}^2 \gamma,$$ as confirmed by numeric simulations in Fig.\ref{fig:Estimators}a, d. As $\gamma \to 0$, the estimator integrates over longer and longer histories and provides an accurate estimation of the phase. 

\subsection{Internal noise}
$K_{\gamma}(t)$ must be generated internally by the estimator during integration. We model the intrinsic unreliability of time-keeping as phase diffusion for $K_{\gamma}$,
\begin{eqnarray}
K_\gamma(t) &=& \Gamma e^{-\gamma t} \sin(\hat\psi_t),\\
d\hat\psi_t &=& \omega dt + \eta(0,\epsilon_{int}) \sqrt{dt}
\end{eqnarray}
where we denote the normalization factor by
$$
\Gamma=\frac{(\epsilon_{int}^2+2\gamma)\sqrt{(\epsilon_{int}^2+2\gamma)^2+16\omega^2}}{4\omega} .
$$

%\newpage
With this, we can write the autocorrelation for noiseless signals $f(t) = \cos \omega t$ and a noisy kernels as, 
\begin{eqnarray}
\langle r(t) r(0) \rangle_{int} &=& \int \int \langle f(t_1) f(t_2) K_{\gamma}(t-t_1) K_{\gamma}(0-t_2) \rangle_{int} \nonumber \\ 
&=& \int \int \cos\omega t_1 \cos\omega t_2 \langle K_{\gamma}(t-t_1) K_{\gamma}(0-t_2) \rangle_{int} \nonumber \\
\end{eqnarray}

%\newpage

Using the definition of the kernels $K_\gamma$, we find,
\begin{widetext}
\begin{eqnarray}
\langle r(t) r(0) \rangle_{int} 
&=& \Gamma^2 \int \int \cos\omega t_1 \cos\omega t_2 e^{-\gamma (t-t_1+t2)} \langle \sin{(\omega (t-t_1)+\psi_{t-t_1})} \sin{(\omega (-t_2)+\psi_{-t_2})} \rangle_{int}  \label{eqn:varr} \\
\langle r(t) \rangle_{int} &=&  \Gamma\int \cos\omega t_1 e^{\gamma (t_1-t)}\langle \sin{(\omega (t-t_1))+\psi(t-t_1)} \rangle_{int}
\label{eqn:meanr}
\end{eqnarray}
\end{widetext}

Note that $\psi_t$ is an unbiased Gaussian random walk started at $\psi_0=0$, and it follows a Normal distribution with variance $\epsilon_{int}^2 t$. Note that if $\theta$ is random number from a Gaussian distribution $N(\mu,\sigma)$, then one has
$$
\langle\cos{\theta}\rangle=e^{-\sigma^2/2}\cos{\mu}
$$
and
$$
\langle\sin{\theta}\rangle=e^{-\sigma^2/2}\sin{\mu}
$$

Using these identities on Eqn.\ref{eqn:varr},\ref{eqn:meanr}, we can compute the variance of the estimator $\sigma^2=\langle r(0)^2\rangle_{int}-\langle r(0)\rangle_{int}^2$ in the leading order of $\epsilon_{int}^2$ as
\begin{equation}
\sigma^2 =  \frac{(2\gamma^4+\gamma^2\omega^2+2 \omega^4) \epsilon^2}{8\gamma\omega^2(\gamma^2+\omega^2)}+O(\epsilon_{int}^4)
\end{equation}
in the regime where $\gamma\ll \omega$ and $\epsilon_{int}^2\ll 1$ we can further simplify the variance to be
\begin{equation}
\sigma^2 \approx \frac{\epsilon_{int}^2}{4\gamma} 
\end{equation}

\begin{figure*}[!htbp]
\begin{center}
\includegraphics[width=\linewidth]{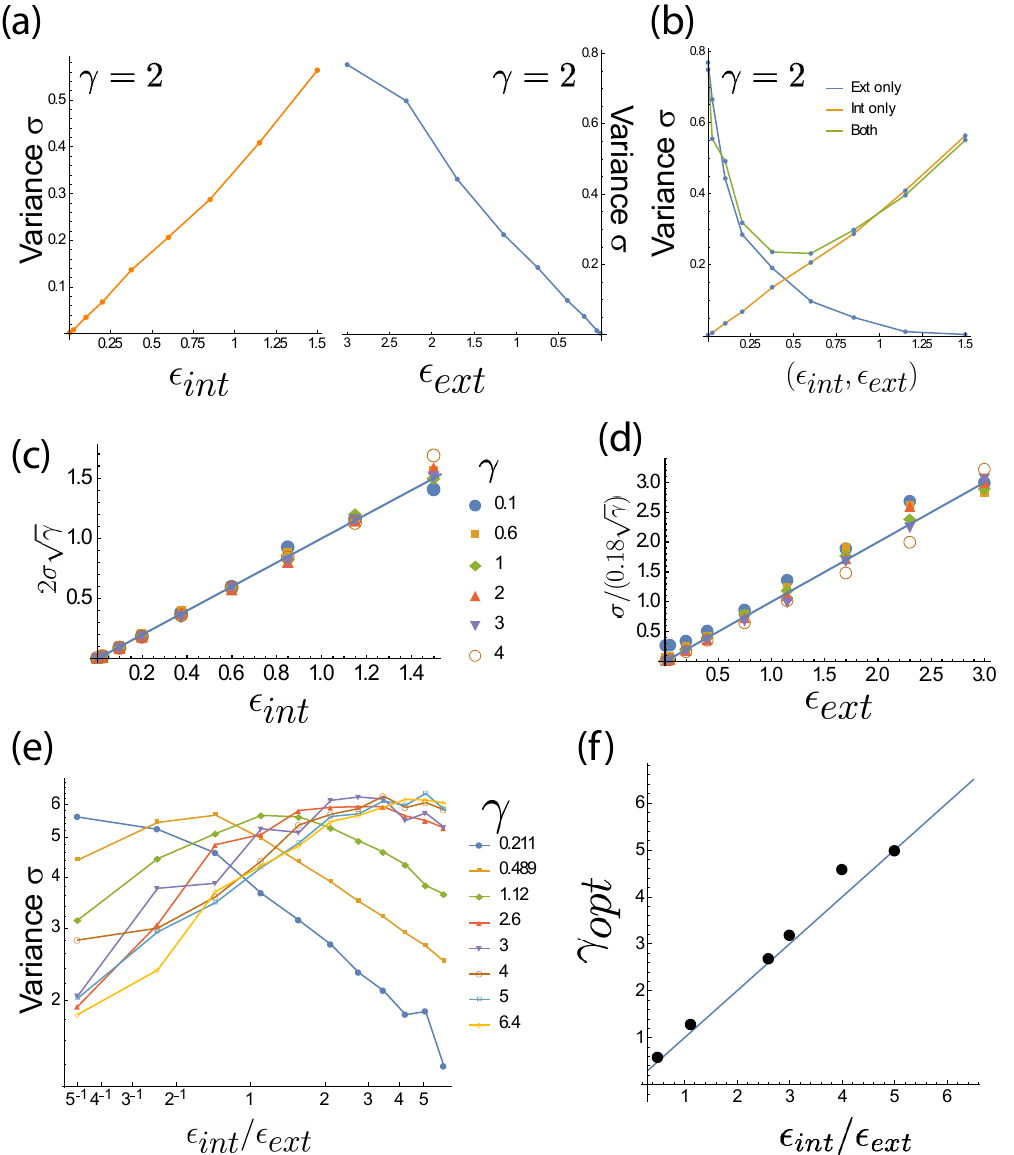}
\caption{Statistical estimators $\hat\phi_\gamma$ with higher precision (i.e., lower variance) with external noise show lower precision with internal noise and vice-versa. (a) For any given $\gamma$, the estimator $\hat\phi_\gamma$  has variance that increases linearly with internal noise $\epsilon_{int}$ and with $\epsilon_{ext}$ (shown here for $\gamma = 2,\omega = 2 \pi$). (b) When both external and internal noise are present at the same time, the resulting variance (green) is approximately the maximum of the variances found if only internal (orange) and only external (blue) noises were present (Here, each point on the x-axis represents a combination of $\epsilon_{int}$ and $\epsilon_{ext}$ , taken in sequence from panel (a). Only $\epsilon_{int}$ is shown on the x-axis. ) (c) With only internal noise, the variance $\sigma^2$ for estimators $\hat\phi_\gamma$ with different $\gamma$ can be collapsed using the formula $\sigma \sim \epsilon_{int}/\sqrt{\gamma}$. (d) Similarly, with only external noise present, the variance $\sigma^2$ satisfies the relationship $\sigma \sim \epsilon_{ext} \sqrt{\gamma}$. (e) Each estimator’s performance peaks at a specific ratio of $\epsilon_{int}/\epsilon_{ext}$ and outperforms other estimators. (f) We find that the optimal estimator has $\gamma_{opt} = \epsilon_{int}/\epsilon_{ext}$.}
\label{fig:Estimators}
\end{center}
\end{figure*}

\textbf{Optimal estimator}
To derive the optimal estimator, note as shown in Fig.\ref{fig:Estimators}c, that with both noises present, the variance is given by $\mbox{max}(\epsilon_{int}^2/\gamma, \epsilon_{ext}^2 \gamma)$. This variance is minimized when the two terms are equal, giving $$\gamma_{opt} \sim {\epsilon_{int}/\epsilon_{ext}}$$.

\textbf{Time-precision trade-off}
With only external noise, we see that slower estimators (i.e. small $\gamma$ leading to longer integration of history) have a higher precision, leading to a trade-off between precision and speed. However, with only internal noise, slower estimators are \textit{less} precise since longer integration times exposes the estimator to more internal noise-related dephasing.  With both kinds of noise present, the optimal estimator $\gamma_{opt} \sim {\epsilon_{int}/\epsilon_{ext}}$, strikes a balance in integration time; integrating any longer would be more negatively affected by internal noise than would be gained by averaging out external noise. Similarly, integrating for less time would insufficiently average out the external noise and not gain as much from lower exposure to internal noise.   The same structure of trade-offs is seen for limit cycle and point attractor-based clocks.

\section{Circle Map - Step Response Curve}

In our main paper, we claim that the variance of the clock state across a population drops $\sigma^2 \Rightarrow \sigma^2/s^2$ at dusk where $s^2 - 1 \sim L/R$ as $L/R \Rightarrow 0$. Data from Langevin simulations was presented. Here we will derive this result using a simple geometric argument about circle maps.

\begin{figure}[!htbp]
\begin{center}
\includegraphics[width=0.8\linewidth]{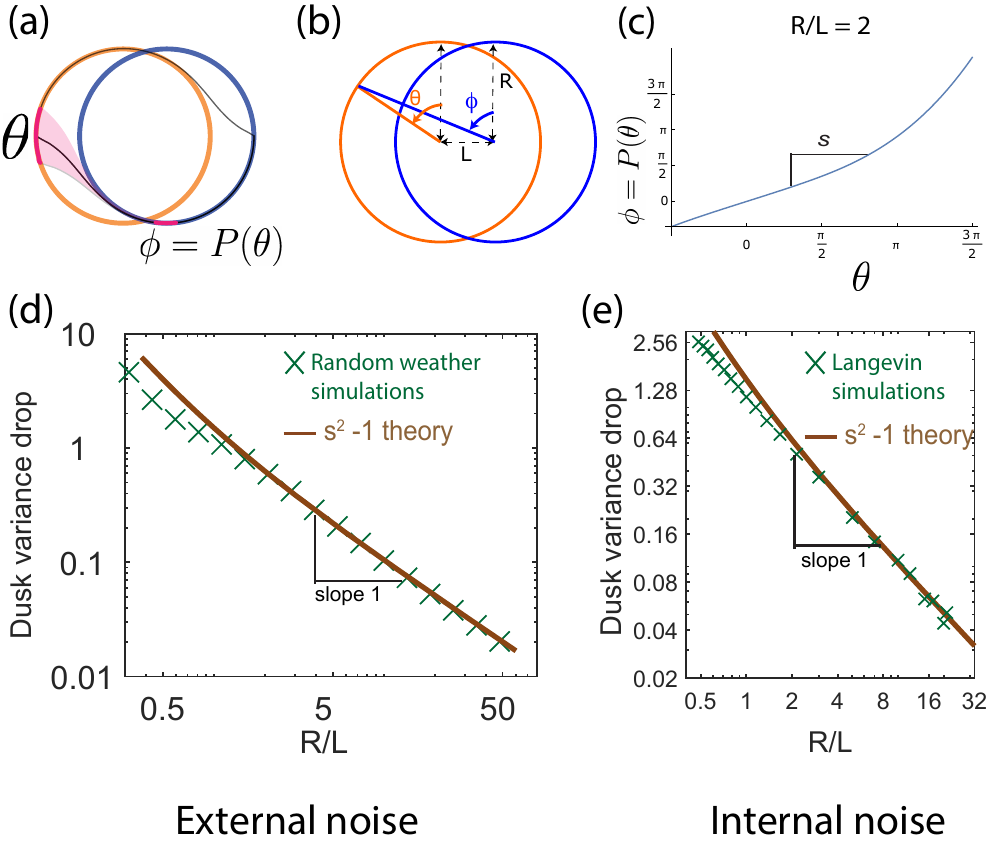}
\caption{The population variance of clock states is reduced by dusk and can be computed geometrically. (a) A population of clocks near state $\theta$ on the day cycle is mapped to the neighborhood of state $\phi$ on the night cycle by the dusk transition.  We define $\phi = P(\theta)$ to be the map relating the clock state $\theta$ on the day cycle just before dusk to its eventual position $\phi$ on the night cycle after dusk (assumed greater than the relaxation time). (b) This map can be analytically computed for circles of size R with centers separated by length L.  (c) For a given R/L = 2 , we obtain $P(\theta)$ shown here. Since $\theta = \pi/2$ corresponds to the dusk time of the entrained trajectory, the slope $s^{-1} = dP/d\theta$ at $\theta = \pi/2$ determines the change in population variance of clock states at dusk. (d,e) The variance drop $s^2$ at dusk, defined as $\sigma^2 \to \sigma^2/s^2$ at dusk, seen in both the external (averaging over weather) and internal noise (averaging over Langevin noise) simulations agree well with the geometrically computed $s(R/L)$, especially at large $R/L$. We find that $s^2 – 1 \sim L/R$ for large-$R/L$ limit cycles.}
\label{fig:phi_is_P_theta__calculation}
\end{center}
\end{figure}

We define $\phi = P_T(\theta)$ to be the phase on the night cycle that a clock evolves to, after time a time $T$, if the lights were suddenly turned off when the clock is at state $\theta$ on the day cycle. See Fig.\ref{fig:phi_is_P_theta__calculation}a,b. In principle, with complex relaxation dynamics between the limit cycles, $P_T(\theta)$ could show complex dependence on $T$. However, we work in a simplified model where the angular frequency of the clock is independent of the amplitude of oscillations. In this limit, $T$ only causes an overall shift in $\phi =P_T(\theta)$; i.e., we can write $P_T(\theta) = P(\theta) + \omega T$ where $\omega$ is the angular frequency of the clock.  In what follows, we will be interested in the derivative of $\partial_\theta P_T(\theta)$; hence we will work with $P(\theta)$ instead of $P_T(\theta)$. 

This circle map, $\phi = P(\theta)$, is important since it determines whether two differing day-time clock states are brought closer or taken further at dusk and thus determines the rate of entrainment of a population to the external signal. Consider two organisms that have nearby but distinct clock states $\theta_0$, $\theta_0+\Delta \theta$ at dusk. After dusk, these two clocks will be mapped to $P(\theta_0)$ and $P(\theta_0 + \Delta \theta) \approx P(\theta_0)  + \Delta \theta dP(\theta)d\theta\vert_{\theta=\theta_0}$ respectively. Thus, dusk changes the  difference between the clock states from $\Delta \theta$ to $\Delta \phi$ where,
\begin{equation}
\Delta \phi \approx \Delta \theta \left. \frac{dP(\theta)}{d\theta}\right\vert_{\theta=\theta_0}
\end{equation}
By a similar argument, if the variance of clock states across a population is $\sigma^2$ before dusk, it will be reduced by,
\begin{equation}
\sigma^2 \xrightarrow{dusk} \sigma^2 \left( \left. \frac{dP(\theta)}{d\theta}\right\vert_{\theta=\theta_0}\right)^2
\end{equation}
This expression is valid in the regime where the population variance $\sigma^2$ is small enough to linearize the circle map $P(\theta)$. Similar considerations apply to the dawn transition between the night and day cycle as well. Both circle maps were recently experimentally characterized for \textit{S. elongatus} in \cite{Leypunskiy2017-al}.

In our simple theoretical model where clock frequency does not change with amplitude (i.e. the radial coordinate), we can easily compute $P(\theta)$ from geometry.
In Fig.\ref{fig:phi_is_P_theta__calculation}, we draw a diagram of the transition from a particle on the day cycle at the phase $\theta$ to the night cycle at the phase $\phi$. By trigonometry, we write
\begin{equation} \phi = P(\theta) = \arctan \left(\frac{L + R \sin \theta}{R \cos \theta}\right), \end{equation}
and derive
\begin{align}
 s^2 - 1 &= \left( \frac{d P(\theta)}{d\theta}\right)^{-2} - 1\\
 &=  \frac{L (2 L^3 + 7 L R^2 - 3 L R^2 \cos(2 \theta) + 
     4 R (2 L^2 + R^2) \sin\theta)}{2 R^2 (R + L \sin\theta)^2}\\
 &=  2 \sin(\theta) \frac{L}{R} + \mathcal{O}\left(  \frac{L}{R} \right)^2,
\end{align}
 where $\theta$ corresponds to the angle on the day cycle at dusk, which is at $\pi/2$ in Fig.\ref{fig:phi_is_P_theta__calculation}a. This equation implies that as the day and night limit cycle gets closer, the geometric focusing effect $s$ converges to one. This asymptotic behavior is intuitive because if $L = 0$, meaning no transition, then the variance should remain the same ($s = 1$, so $\sigma^2 \rightarrow \sigma^2/1^2$ at the transition).

Remarkably, our geometric derivation of $s^2 - 1$ matches the variance drop $\sigma^2 \to \sigma^2 /s^2$ seen in stochastic simulations of weather conditions; see Fig.\ref{fig:phi_is_P_theta__calculation}d. The variance gain during the day is the result of the fluctuation of sunlight, simulated as random dark pulses of random intervals, amplitude and time of delivery. Such variance is accumulated during the day and the drop over dusk time is measured (green Xs). 

Fig.\ref{fig:phi_is_P_theta__calculation}e shows the variance drop seen in simulations with internal noise in Langevin simulations. While the cause of variance increase during the day is different (finite copy number effects), the variance drop at dusk agrees well with the geometric computation of $s^2$ and thus with the external noise simulations as well. In both cases, the simulations and geometric theory show that $s^2 - 1 \sim L/R$ as $L/R \Rightarrow 0$.

\begin{figure}[!htbp]
\begin{center}
\includegraphics[width=\linewidth]{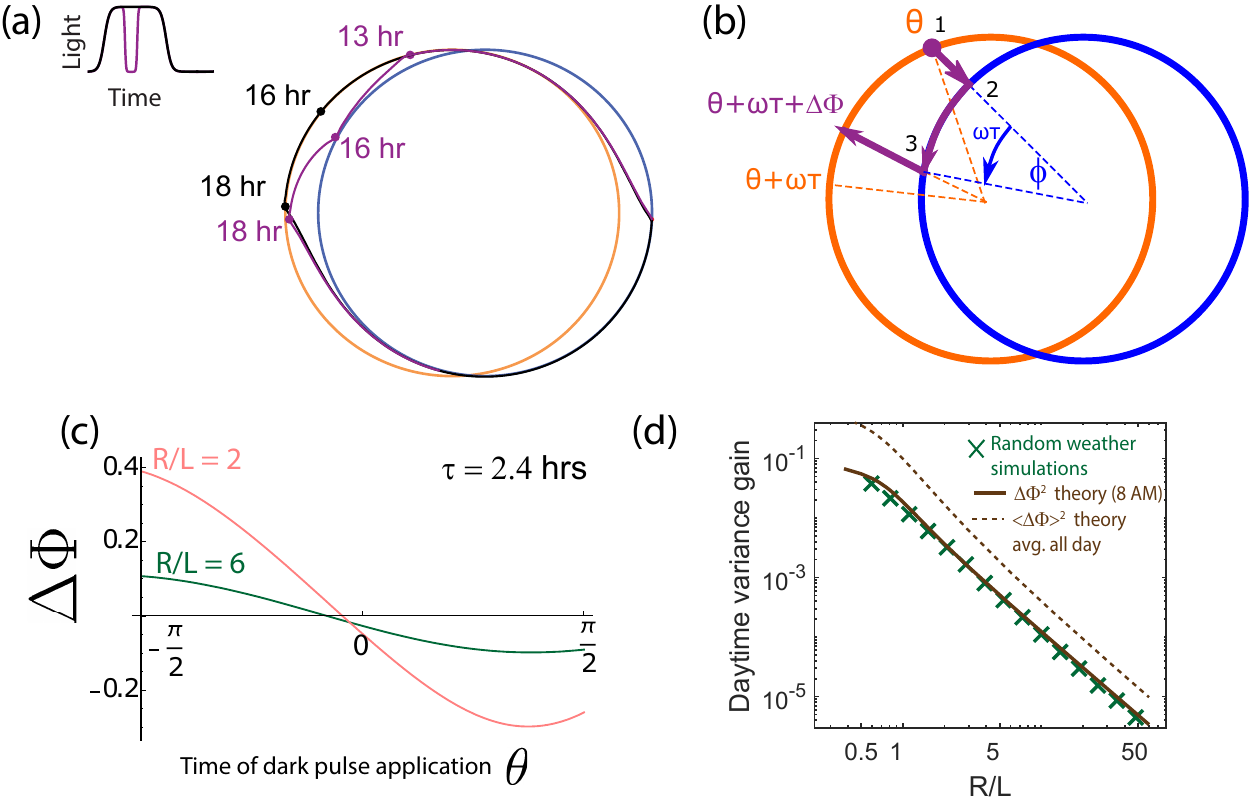}
\caption{Increase in population variance due to random weather conditions can be estimated from the phase shifts $\Delta \Phi$ due to dark pulses (i.e., the Phase Response Curve). (a) A single dark pulse administered during the day shifts the phase of a clock (purple) relative to a clock that experiences no such dark pulse (black). (b) We can compute the phase shift $\Delta \Phi$ due to such a dark pulse geometrically by computing the deviation in trajectory. Assuming a dark pulse of length $\tau$, the clock evolves for a time $\tau$ according to the night cycle dynamics. At the end of such a pulse, we switch back to the day limit cycle and compute the resulting phase shift $\Delta \Phi$.  (c) The resulting phase shift $\Delta \Phi$ due to a pulse of length $\tau = 2.4$ hrs, depends on the time $\theta$ when it is administered but is generally smaller for larger $R/L$. (d) We find that $\Delta \Phi^2$ for a specific $\tau = 2.4$ hrs dark pulse administered at the same time (8 AM) falls as $(L/R)^2$ for large-$R/L$ limit cycles. This trend matches the variance gain $\sigma^2_{clouds}$ seen in stochastic simulations that average over random weather conditions (pulses of different length, intensity and time of application). The broken brown curve shows a theoretical prediction for such an average $\langle\Delta \Phi^2\rangle$, obtained by sampling the curve shown in (c) at different points of application and differing intensity. Despite the presence of a variance-reducing zero around mid-day in (c), $\sigma^2_{clouds}$ drops as $(L/R)^2$, much as $\Delta \Phi^2$ for any particular pulse. (Brown theory curves translated together using one fitting parameter.).}
\label{fig:SI_Dark_Pulse}
\end{center}
\end{figure}

\section{Dark pulse phase shift - Phase response curve}
During the daytime, sunlight intensity fluctuates because of cloud cover and we have referred to these fluctuations as external noise. In our simulations, we subject each individual in a population to a different realization of these weather conditions and compute the resulting population variation of clock state. Such variation limits the ability of the cell to read out the objective time from the clock state.

Here, we relate the population variance caused by random cloud cover to the geometrically computed Phase Response Curve (PRC) due to a single dark pulse administered during the day. Using this geometric method, we will find that the ability of limit cycle to withstand external intensity fluctuations increases with $R/L$, the size $R$ of limit cycles relative to their separation $L$. In particular, we will show geometrically that the variance gain during the day $\sigma^2 \Rightarrow \sigma^2 + \sigma^2_{clouds}$ scales as $(L/R)^2$, in perfect agree with stochastic weather simulations.

%\subsection{Effect on limit cycles}

To compute the scaling relationship of $ \sigma^2_{clouds}$, we compute the phase shift $\Delta \Phi$ caused by a single dark pulse with width $\tau$ on the limit cycles with angular speed $\omega$ (i.e., the Phase Response Curve (PRC) corresponding to such a dark pulse). Fig.\ref{fig:SI_Dark_Pulse}a shows an example of a dark pulse in the signal and how it affects the trajectory. Consider a clock at state $\theta$ on the day cycle. A dark pulse of length $\tau$ administered just then will change the dynamics to that of the night cycle. This clock has state $\phi = P(\theta)$ with respect to the night cycle and will evolve for a time $\tau$ according to the night cycle dynamics, reaching a new state $\phi+\omega \tau$, at a radial position determined by $R,L$. At the end of the dark pulse, we use the night-day circle map, $\theta = Q(\phi)$, to find the clock state back on the day cycle. Note that all these shifts depend on the limit cycle geometry, i.e., on $R$ and $L$, as shown in Fig.\ref{fig:SI_Dark_Pulse}.
Similar to how we compute the mapping in the previous section, we can write each mapping using simple trigonometry:

\begin{equation}  \phi = P(\theta) = \arctan \left(\frac{L+R \sin \theta}{R \cos \theta}\right)
\end{equation}
and
\begin{equation} \theta^* = Q(\phi) = \arctan \left(\frac{-L + R \sin(\phi + \omega \tau)}{ R \cos (\phi + \omega \tau)}\right).\end{equation}

Notice the mapping $Q$ only differs from $P$ by changing $L$ to $-L$. We also include the diagram showing the transition due to dark pulse in Fig.\ref{fig:SI_Dark_Pulse}. The process ``1'' corresponds to $\phi = P(\theta)$, ``2'' corresponds to the rotation on the night cycle $\phi \rightarrow \phi + \omega \tau$, and ``3'' corresponds to the transition back to the day cycle $\theta^* = Q(\phi+\omega \tau)$. Combining this 3 processes, we write $\theta^*$ as $\theta^*(\theta, \tau, L/R)$ and expand it in the limit that $L/R\Rightarrow 0$ to obtain that

\begin{equation}
\Delta \Phi = -\frac{L}{R} \left(\cos(\theta + \omega \tau) - \cos(\theta)\right) + \mathcal{O}\left(\frac{L}{R}\right)^2
\end{equation}
where $\Delta \Phi = \theta^* - (\theta + \omega\tau)$ because $\theta + \omega \tau$ is the phase of the clock if it did not experience the dark pulse.

This expression $\Delta \Phi$ indicates the amount of phase shifted that the cloud causes. With different clocks experiencing different weather conditions, the variance gained among the population due to the fluctuation of sunlight grows like $|\Delta \Phi|^2 \sim (L/R)^2$. We see good agreement between stochastic weather simulations and this geometric computation as shown in Fig.\ref{fig:SI_Dark_Pulse}d.

In this calculation, we focused on dark pulses administered at a fixed generic time (8 AM in Fig.\ref{fig:SI_Dark_Pulse}d). However, the PRC $\Delta \Phi(\theta)$ for dark pulses has a zero at a specific time of the day (see Fig.\ref{fig:SI_Dark_Pulse}c). That is, for each dark pulse of width $\tau$, there exists a time of administration such that $\Delta \Phi = 0$! In fact, such a dark pulse has an entraining effect, reducing the population variance.  Such an effect is seen in Fig.3c, where the population variance drops in the middle of the day. We leave experimental and theoretical investigation of the counter-intuitive effects of such specially time dark pulses to future work.

Here, we show that even if we include such dark pulses with an entraining effect, the variance gained at the end of the day is still proportional to $(L/R)^2$ in the limit that $L/R$ goes to zero. To simplify our derivation but retain the essence of what dark pulses do during the daytime, let's us consider dark pulses coming at three times: in the morning ($\theta = -\pi/2$),  around noon ($\theta = -\omega \tau/2$ with small $\omega \tau$), and in the evening ($\theta = \pi/2$). Starting the day with variance $\sigma_0^2$, by the end of the day the variance becomes
\begin{align}
\sigma^2 &= \frac{\sigma_0^2 + (\Delta \Phi)^2_{\theta = -\pi/2}}{(1+ \left( \frac{d \Delta\Phi}{d\theta}\right)_{\theta = -\omega \tau/2})^2} + (\Delta \Phi)^2_{\theta = \pi/2}\\
&\approx \frac{\sigma_0^2 + \left(\frac{L}{R} \sin \omega \tau \right)^2}{\left(1+\frac{2L}{R} \sin \left(\frac{\omega \tau}{2}\right)\right)^2} + \left(\frac{L}{R} \sin \omega \tau \right)^2\\
\sigma^2 &\approx \sigma_0^2 + 2 \left(\frac{L}{R} \sin \omega \tau\right)^2 + \mathcal{O}\left(\frac{L}{R}\right)^2.
\end{align} Thus, the variance gained due to fluctuation, $\sigma^2-\sigma_0^2 = \sigma^2_{clouds}$, is proportional to $(L/R)^2$. This simple derivation may not rigorously reflect the correct constant in front of $(L/R)^2$ term, but the full rigorous derivation, concerning the dark pulses coming randomly at random time during the day, should yield the same power law dependent on $L/R$. Fig.\ref{fig:SI_Dark_Pulse}d shows that averaging $\Delta \Phi^2$ over pulses administered at different times numerically (dashed line) results in the same power law as for single pulses and as seen in stochastic weather simulations.

\section{Langevin model of finite copy number fluctuations} 
Chemical reactions that occur in the bulk of a homogeneous solution can be described by a set of ordinary differential equations. However, within a single cell the copy number of molecule is limited and thus the reaction carries internal noise from the stochastic fluctuations. Gillespie showed that chemical reactions under finite copy number can be approximated by a Langevin dynamics using the following argument \cite{Gillespie2007-yr},

Consider an elementary reaction 
\begin{equation}
~~~~~~~~~\ce{A + B -> C + D} 
\end{equation}
with the forward rate constant $k_{+}$, during each infinitesimal time $\delta t$, the probability of the occurrence of this reaction follows a Poisson distribution whose mean and variance both equals to $R_{+} \delta t= k_{+}\cdot N_A\cdot N_B\cdot \delta t$.
Integration over a larger time step, the Poisson distribution can be approximated into a Gaussian form, resulting in Langevin dynamics,
\begin{equation}
{\rm d} N_a = - k_{+}\cdot N_A\cdot N_B \cdot {\rm d} t + \sqrt{R_{+}} {\rm d} W
\end{equation}
where $W$ is a standard Wiener process of mean $0$ and autocorrelation function $\langle W(t_1)W(t_2)\rangle=\delta(t_1-t_2)$.

To describe a chemical reaction network, the Langevin equation for each species consists of contributions to the noise from each reaction where the species is involved. Now consider adding another reaction 
\begin{equation}
~~~~~~~~~\ce{C + D -> A + B} 
\end{equation}
with the rate constant $k_-$, then the Langevin equation for species A becomes,
\begin{equation}
{\rm d} N_a = - k_{+}\cdot N_A\cdot N_B  \cdot {\rm d} t + k_-\cdot N_C\cdot N_D  \cdot {\rm d} t + \sqrt{R_{+}} {\rm d} W_1 + \sqrt{R_-} {\rm d} W_2
\end{equation}
where $R_+=k_{+}\cdot N_A\cdot N_B$ and $R_-=k_{-}\cdot N_C\cdot N_D$ respectively denote the number rates of the forward and the backward reaction; ${\rm d} W_1$ and ${\rm d} W_2$ are identical independent standard Wiener processes. 

To fully determine the effect of the noise using the Langevin dynamics for a chemical reaction network, one needs to consider all of the reactions corresponding to the species of interest; the noise term usually becomes time-dependent and multiplicative. To simplify the description of internal noise in our phenomenological model of limit cycle/ point attractor, we take a first order approximation that the diffusion coefficient in the reaction coordinate space is homogeneous in both space and time. (See similar treatments of another biological system in \cite{Potoyan2014-so}. In contrast, our explicit KaiABC simulations, presented later, do not make this simplifying assumption of homogeneous diffusion.) This allows us to write a 2-dimension phenomenological stochastic differential equation
\begin{equation}
{\rm d} \vec z = f(\vec z, t)\cdot {\rm d} t + \sqrt{2 D} \cdot {\rm d} \vec W
\end{equation}
where the $f(\vec z, t)$ denotes the deterministic dynamics driven by day-night cycles and the diffusion constant is assumed to be proportional to the total number of Kai-C molecules within the cell.

\subsection{Population variance}
For the cell to carry out a reliable computation, the population variance from the internal noise needs to be reduced. Such noise reduction comes from the dynamics of the attractor. In the limit cycle attractor mechanism, the internal noise reduction is performed only along the radial axis but not along the flat attractor direction. 

In contrast, the point attractor mechanism is able to limit population variance due to internal noise in all directions due to the effective `curvature' of the dynamics.  Here we analytically estimate the steady-state population variance for a point attractor mechanism. The population variance is together determined by the diffusive term $ \sqrt{2 D} \cdot {\rm d} \vec W$, and the noise reduction effect from the restoring force of the point attractor's harmonic well. During each infinitesimal time $\delta t$, the internal noise increase the variance by
\begin{equation}
\sigma ^2(t+\delta t) = \sigma^2(t)+ 2 D\delta t.
\end{equation}
In contrast, the overdamped deterministic motion within a harmonic well provides a focusing effect that reduces the variance exponentially with time. To quantify this focusing effect, consider a 1-d overdamped dynamics of a particle within a harmonic energy well of $V(r)=k\cdot r^2$. The solution to the equation of motion is $r(t)=r_0\cdot e^{-2kt}$, with initial position $r(0)=r_0$. Consider an ensemble of points with a mean initial position $\mu_0$ and a initial variance of $\sigma ^2_0$, one can solve the dynamics of the mean as
\begin{equation}
\mu(t)=\mu_0\cdot e^{-2kt}
\end{equation}
and the dynamics of the variance as
\begin{equation}
\sigma^2(t)=\sigma^2_0\cdot e^{-4kt}
\end{equation}
Thus, per infinitesimal time $\delta t$, the geometric focusing effect of the energy well of the point attractor reduces the population variance by 
\begin{equation}
\sigma^2(t+\delta t)=\sigma^2(t)/g 
\end{equation}
where $g=e^{4k\delta t}$.

Under the competition between the spreading effect from the internal noise and the geometrical focusing effect from the deterministic dynamics, the population variance reaches a steady value solved by 
\begin{equation}
\sigma^2_{st}=\frac{\sigma^2_{st}+ 2 D\delta t}{g} =\frac{\sigma^2_{st}+ 2 D\delta t}{e^{4k\delta t}}
\end{equation}
and by taking the limit of $\delta t$ goes to $0$, we have $\sigma^2_{st}=D/2k$.

\begin{figure*}[!htbp]
\begin{center}
\includegraphics[width=\linewidth]{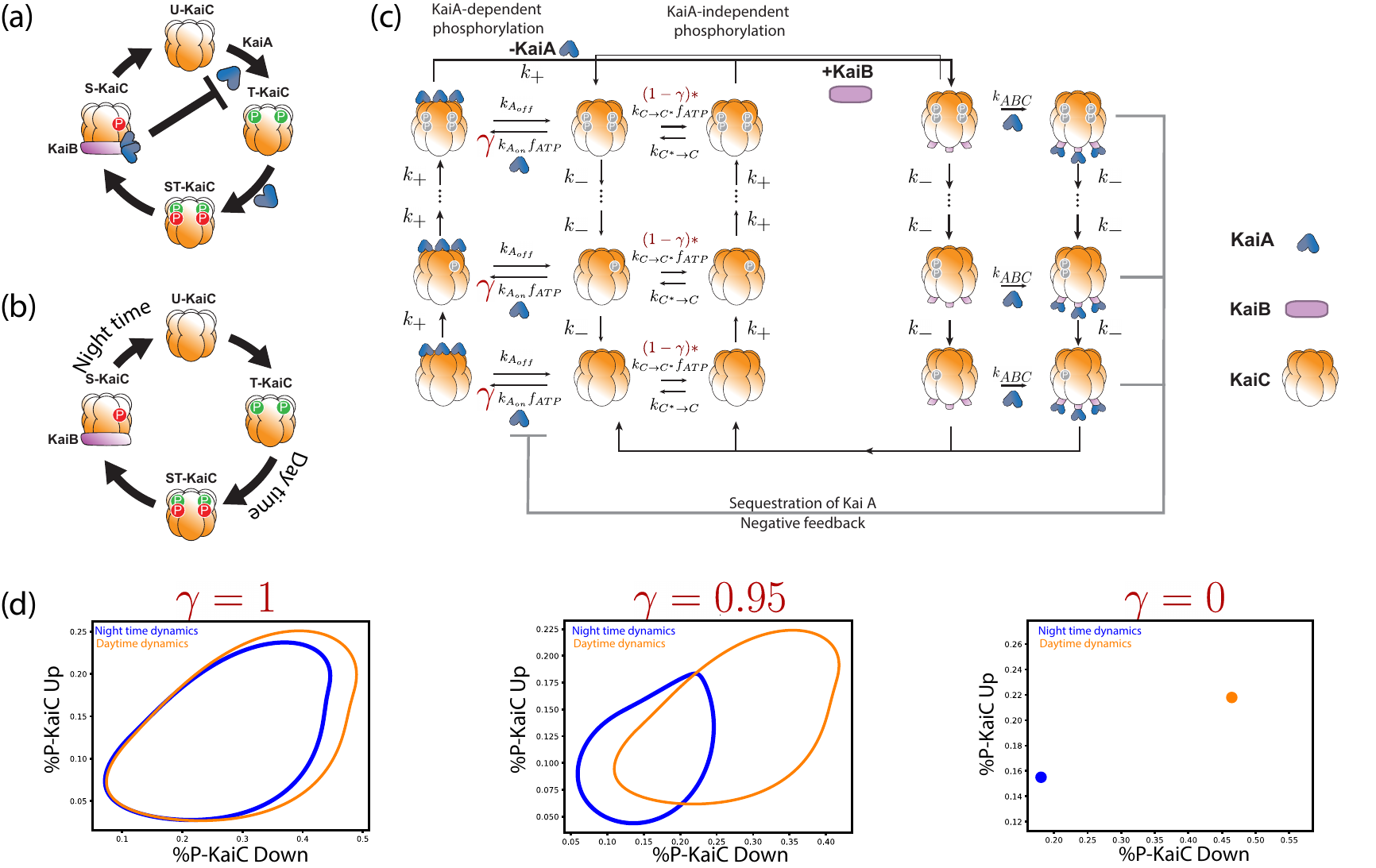}
\caption{Explicit biochemical KaiABC model simulated using the Gillespie algorithm. (a) The experimentally well-characterized clock in \textit{S. elongatus} consists of a negative feedback-enabled self-sutained oscillator.  KaiBC complexes sequester KaiA, preventing runaway KaiC molecules from going through the cycle independently. (b) The genome of \textit{P. marinus} lacks \textit{kaiA}.  We assume a minimal model consistent with known facts \cite{Rust2007-op} about this clock; KaiC phosphorylation proceeds without KaiA and hence different KaiC hexamers can proceed independently through the cycle.  
(c) We combine both clocks in one model with an interpolating parameter $\gamma$ that selects between an \textit{S. elongatus}-like KaiA-dependent pathway and an \textit{P. marins}-like KaiA-independent pathway. All reactions shown are assumed to be first order mass-action kinetics. We simulate such a system at different overall copy numbers $N$ using the Gillespie algorithm. (d) We find limit cycles for $\gamma > 0.9$. The resulting limit cycles for $\gamma=1,0.95$ violate the simplifying assumptions used in our dynamical systems (e.g., non-circular cycles of different size); and yet our  results are qualitatively validated by this model (Fig.4).}
\label{fig:KaiABCmodel}
\end{center}
\end{figure*}

\section{Explicit KaiABC biomolecular model}
We derived our results in two distinct ways: (a) using an abstract theory of estimators, (b) using a simplified dynamical systems picture of circadian clocks. Here we illustrate our results in a third independent way - using Gillespie simulations of an explicit biomolecular KaiABC model. This model, based on recent experiments, violates the simplifying assumptions and idealizations made earlier - such as assuming circular limit cycles of the same size during the day and night, Langevin approximation of internal noise with a homogeneous and time-independent diffusion coefficient, the two dimensional nature of the dynamical systems etc. Nevertheless, we find qualitatively similar results, showing that our results rely only on the essential properties of these systems such as the existence of a continuous attractor.

\subsection{\textit{S . elongatus} clock - hexamers with collective KaiA feedback}
The \textit{S. elongatus} clock has been well-characterized experimentally \cite{Bryant2003-zw,Gutu2013-oy, Dufresne2003-gh, Kitayama2003-na} - see Fig.\ref{fig:KaiABCmodel}a. The clock is fundamentally based on the ordered phosphorylation and dephosphorylation of KaiC \cite{Rust2007-op}. Phosphorylation of KaiC is KaiA-dependent which allows for feedback that enables collective coherent oscillations in a cell. After complete phosphorylation of KaiA-C complexes (usually by the end of the day), KaiC forms a KaiB-C complex which then dephosphorylates in an ordered manner. Crucially, the KaiB-C complex also sequesters KaiA in a KaiABC complex, reducing the pool of available KaiA for phosphorylation of other KaiC hexamers. This negative feedback enables coherent oscillations of the population of KaiC molcules in a single cell\cite{Rust2007-op}. 

\subsection{\textit{P. marinus} model - independent hexamers}
\textit{P. marinus} lacks the \textit{kaiA} gene but possesses and expresses \textit{kaiB} and \textit{kaiC}. While the details of the protein clock are not fully known, gene expression shows cycling in cycling conditions but decays in constant conditions \cite{Holtzendorff2008-bj}. A conservative model, consistent with all these known facts about \textit{P. marinus}, is shown in Fig.\ref{fig:KaiABCmodel}b; without KaiA feedback, different hexamer units phosphorylate independently and settle to a hyperphosphorylated state at the end of the day. At night, they dephosphorylate along a distinct pathway (homologous to that used by \textit{S. elongatus} but without KaiA) and reach a hypophosphorylated state by dawn. 

\textbf{Hybrid model}
We created the following hybrid model that includes \textit{S. elongatus} and \textit{P. marinus} models as different limits. In our model, shown in Fig.\ref{fig:KaiABCmodel}c, KaiC has a KaiA-dependent phosphorylation pathway, much like in \textit{S. elongatus}, that is used during the day and driven forward by ATP. 

But to also include \textit{P. marinus}-like behavior in the model, we allow for a second parallel phosphorylation pathway for KaiC that is independent of KaiA. The relative access of these two pathways is controlled by a parameter $\gamma$. When $\gamma = 1$, only the \textit{S. elongatus}-like KaiA dependent pathway is accessible. When $\gamma = 0$, only the \textit{P. marinus}-like KaiA independent pathway is accessible. Collectively, we call these states along these phosphorylation pathways, the UP states of KaiC - phosphorylation are going UP along these pathways which are usually used during the day.

After maximum phosphorylation (usually at dusk), KaiA unbinds (if present) and a KaiB-based dephosphorylation pathway takes over (common to both systems). We call these states the DOWN states of KaiC.

Critically, KaiA is assumed to be sequestered through the formation of KaiABC complexes during this dephosphorylation stage. In \textit{S. elongatus}, reduced KaiA availability prevents other KaiC hexamers from proceeding independently through the UP stage while most of the population is in the DOWN state. Such negative feedback is critical in maintaining free-running limit cycle oscillations in \textit{S. elongatus}. 

However, as $\gamma \to 0$, the KaiA-independent pathway is more active and thus the system effectively has no feedback. In fact, we find that at about $\gamma \approx 0.82$, sustained oscillations disappear (for kinetic parameters used here and reported below). Hence we chose $\gamma = 1, 0.95, 0$ as representative of two limit cycle-based and one point-attractor based clock respectively.

\subsection{Gillespie simulations}
We ran explicit Gillespie simulations corresponding to the deterministic equations above at different overall copy number $N$ with fixed stoichiometric ratios of the molecules KaiA,B,C. 

We simulated external input noise by varying the ATP levels during the day. External noise in these simulations were implemented by changing ATP levels in the following way: we fluctuated the ATP levels $f_{ATP} = ATP/(ATP+ADP)$ during the day between the $f_{ATP}^{day}$ and $f_{ATP}^{night} + (f_{ATP}^{day}- f_{ATP}^{night})/3$, where $f_{ATP}^{day}$, $f_{ATP}^{night}$ are the ATP values during a cloudless day and night respectively. We used the day and night ATP levels for different $\gamma$ that ensure that the limit cycles had periods comparable to $24$ hours. For $\gamma = 1$, we used ATP/ADP ratios of $f_{ATP}^{day} = 0.55,f_{ATP}^{night}= 0.45$. For $\gamma = 0.95$, we used $f_{ATP}^{day} =0.57,f_{ATP}^{night} =0.17$ and for $\gamma = 0$, $f_{ATP}^{day} =0.8,f_{ATP}^{night} =0.2$. The corresponding limit cycles and point attractors are shown in Fig.\ref{fig:KaiABCmodel}d.

We used the following kinetic parameters in all simulations:
$dt = 0.01 \; hr, k_{+} = k_{-} = 2m \cdot 0.04932\; hr^{-1}, k_{Aon} = 0.2466\; \mu M^{-1} hr^{-1}, k_{Aoff} = 0.02466\; hr^{-1}, k_{C\to C*} = 0.2466\; hr^{-1}, k_{C* \to C} = 0.1 \;k_{C \to C*}, k_{ABC} = 123.30 \;hr^{-1}, m = 18$. We set up Kai C and Kai A in a $1:1$ stoichiometric ratio, each present at a copy number $N$ where $N$ was varied as shown in Fig.5c.  
These rates are consistent with those measured in \cite{Quinn2001-um,Lovell1992-sf}.

Much like with Langevin simulations of dynamical systems, we run the Gillespie simulation until equilibration of the population. However, the system appears to reach the equilibrium state much faster (only over 5 light-dark cycles of 12h:12h). We extracted one day of such a trajectory on day 6 and repeated the simulation 100-400 times. We repeat 400 times when the copy number is low ($< 1200$) since the spread will be big and we found that the probability distribution is not smooth. We run only 100 times for the high copy number ($> 1200$). Pooling together these trajectories, we computed the mutual information between clock statea (i.e., $(u,d)$ where $u$ is the net phosphorylation state of KaiC in the up-pathways and $d$ is the net phosphorylation state of KaiC in the KaiB-bound `down' pathways in Fig.\ref{fig:KaiABCmodel}c ) and time of day. The $(u,d)$ space was binned using bins of fixed size of dimension $(0.05, 0.05)$ while the 24 hr time-of-day was binned with bins of size $0.5$ hrs.

\subsection{Violation of simplifying assumptions}
With these choices of $\gamma$, we see in Fig.\ref{fig:KaiABCmodel}d, that this model has limit cycles of different size during the day and night; these cycles are not circular in any projection.  Further, the relaxation time between attractors varies with $\gamma$ and in general, differs from $\tau_{relax}$ used in the simulation of limit cycle attractors in the paper. While we assumed a time- and state-independent diffusion constant to model internal noise in the dynamical system, the strength of fluctuations in the explicit KaiABC model can vary with time, as KaiA is sequestered and released by KaiBC over the course of the day-night cycle.

Thus this model violates the simplifying assumptions made in the dynamical systems model. Despite such violations, this explicit KaiABC biomolecular model qualitatively reproduces our dynamical systems-based results since the latter only rely an elementary coarse feature of the system - the existence of a flat attractor direction that can project out external noise but is then susceptible to internal noise.

\section{Supplementary Methods}

\subsection{Dynamical system - Simulation details}

We simulate two kinds of dynamical systems in this paper; limit cycles and point attractors. In each case, we simulate a population of clocks, each represented by a particle in the given dynamical system, subject to external and/or internal noise.

The equation that we use for the simulation is
\begin{align}
\frac{d r}{d t} &= \alpha r - | \alpha | \frac{r^3}{R^2}\\
\frac{d\theta}{d t} &= \omega
\end{align}
where $|\alpha| = 1/\tau_\text{relax}$. We use $\alpha = 5$ for limit cycle system and $\alpha = -5$ for point attractor system. For limit cycles, $R$ controls the size of the attractor. For point attractors, we set $R = 1000 L$, where $L$ is the separation of the day and night attractor. In such a limit, the point attractors are quadratic potentials with linear restoring forces since $\frac{r^3}{R^2}$ is small. The center of the cycle and point attractors during the day are assumed to be at $(-L,0)$ and at $(0,0)$ at night.

We evolve our dynamical system using the Fourth Order Runge-Kutta method with time step $dt = 0.001$ days until the value of mutual information from one day to the next does not change by more than 2-3\% - i.e,. the  system has reached steady state. Reaching steady-state usually takes around 200 days, but if the ratio of $L/R$ is smaller than $0.1$, then we may need to run the simulation until day 500 to reach an equilibrium (See speed-error tradeoff in Fig.5). 

For limit cycles, we initialize the population of $10^4$ particles by uniformly distributing them along the perimeter of the night cycle. In the point attractor system, we initialize a population of $10^5$ at the night-time point attractor.

We use a larger population with point attractors since the particles tend to be distributed over a larger area of the dynamical system. Note that we bin the population by position to compute mututal information between position in the 2d state space and time. Doing so reliably requires a smooth distribution after binning. For limit cycles, the particles usually stay close to attractor and thus provide sufficient count in each bin. However, for the point attractor, the population is usually spread over the entire 2d area between the two point attractors. Therefore, we need $10^5$ particles to get an accurate value of mutual information of point attractor system.

\textbf{External signal and weather fluctuations}
We generate a square wave of period $24$ hours to model the day-night cycle of light on Earth with the day length of 12 hours.  However, such a square wave is modulated by weather fluctuations, e.g. periods of reduced intensity due to passing clouds during the daytime. We model such fluctuating intensity as follows. We assume each weather condition lasts a random interval of time drawn from an exponential distribution of mean $2.4$ hrs (1/10 of a day). During a given weather condition, we set the intensity of light to a random value, drawn uniformly from $[0,1]$ where $1$ represents the maximum intensity during the day. (At night, the intensity is held at zero with no fluctuations.) 

When the light intensity is reduced during the day to a value $\rho \in [0,1]$, we switch the dynamics to an alternative limit cycle (or point attractor) at a fractional distance $\rho$ between the ideal day and night cycles.  For example, assume the night cycle is centered at $(0, 0)$ and the day cycle is centered at $(-L, 0)$. During a weather condition with intensity $\rho \in [0,1]$, we follow dynamics due to a limit cycle located at $(-\rho L, 0)$.  We follow the same rules for the point attractor. In both cases, the switches of dynamics in response to the changing weather is instantaneous, though the clock states itself is continuous and responds at a finite rate to an instantaneous switch in dynamics. Each individual particle is subject to a different realization of the weather conditions described above.

\textbf{Internal noise}
The internal noise represents any source of stochasticity intrinsic to a single cell that would exist even in constant conditions. Such noise could be due to finite copy numbers of molecules, bursty of transcription etc. We model such internal noise by adding Langevin noise to the dynamical equations as described in the section on Langevin noise. Each individual particle in our simulation is subject to a independent random realizations of such Langevin noise. We then bin the population and compute mutual information by the same procedure as for external signals above.

\subsection{Measures of clock time-telling quality}
We develop and use two distinct measures of performance of noisy clocks driven by noisy inputs. 

\textbf{Mutual information:} The performance of the clock is quantified by the mutual information between the clock state $\vec{c}$ and the time $t$,
\begin{equation}
MI(C; T) = \sum_{\vec{c}\in C, t \in T} p(\vec{c}, t) \log_2\left(\frac{ p(\vec{c}, t)}{p(\vec{c})p(t)}\right)
\end{equation} for all $\vec{c}$ in the set of available positions $C$ and all $t$ in the available time bins $T$. (In the dynamical systems model, $\vec{c}$ represents the position in the 2d $r,t$ plane. For the explicit KaiABC biomolecular model, $\vec{c}$ represents the phosophorylation state of KaiC.) 
We simulate a population of clocks, where each clock is subject to a different realization of input signals, representing different weather conditions and also subject to different realizations of internal Langevin noise (or Gillespie fluctuations). We then collect the trajectories of each clock on the last day of the simulations and calculate the probability distribution $p(\vec{c}\vert t)$ of clock states at a given (objective) time $t \in [0,24]$ hrs of the last day in the simulation. The probability function $p(\vec{c})$ is calculated by accumulating the distribution of $p(\vec{c}\vert t)$ over time $t \in [0, 24]$ hrs of the last day. The position $\vec{c}$ and time $t$ are binned into different bins depending on their values. We start the minimum and maximum values of the bins to the minimum and maximum values of the variables. The bin size in the time dimension is 0.48 hrs or 28.8 minutes, while The bin size in the x and y dimensions are both $0.01$.

We refer to this mutual information measure as `Precision' in Fig.1b, 3f, 5a,5c.

\textbf{Population variance along direction of motion:} Mutual information is a good indicative of how well the clock encodes information about time. However, it is calculated for the entire day. Often, we want to see how the time-telling ability of a clock changes during the day (e.g., day vs night or before and after dusk). Hence we develop a new measure, closely related to mutual information, but can be computed at specific times of day.

Intuitively, the mutual information quantifies how much the population distributions of clock states at different times $t$ overlap. If these distributions are not overlapping, the clock state is a good readout of the time $t$. Such distributions are shown in Fig.3b and 4b (purple). 

We argue that only the spread of the clock distribution along the direction of motion of the clock in state space affects mutual information. The spread of the distribution in orthogonal directions does not affect mutual information as much.

\begin{figure}
\begin{center}
\includegraphics[width=0.7\linewidth]{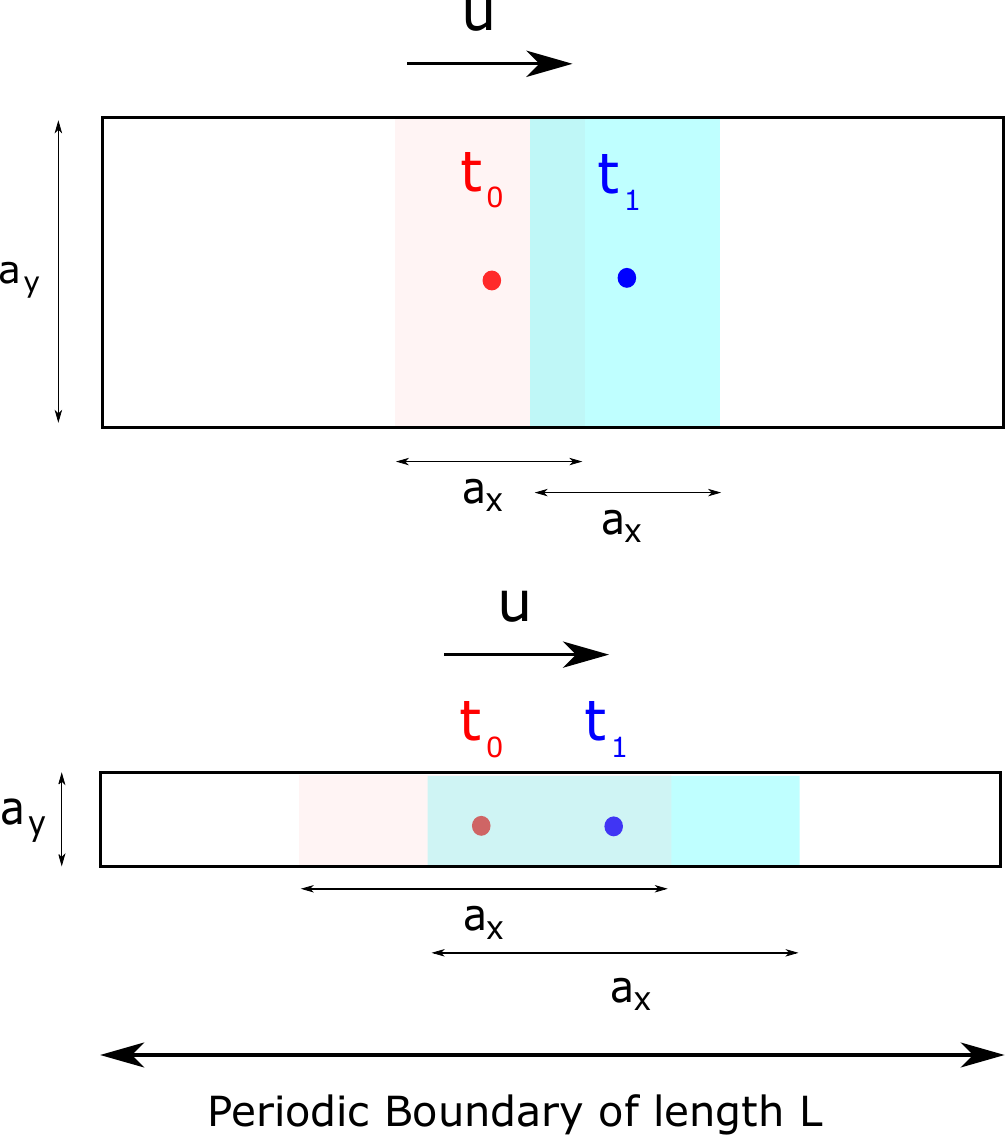}
\caption{Mutual information $MI(\vec{c},t)$ between clock state $\vec{c}$ and time $t$ is only affected by the variance of the clock state distribution $p(\vec{c}\vert t)$ at a given time $t$ along the direction of motion and not orthogonal to it.  In this toy example, we assume the distribution $p(\vec{c}\vert t)$ to be supported on a rectangle of size $a_x$ and $a_y$ in a 2d clock state space. The clock state moves at a speed $u$ in the x-direction. Time telling quality is affected by how much the population at different times overlap with each other. Consequently, clocks with large $a_x$ and small $a_y$ (\textit{bottom}) have lower mutual information $MI(\vec{c},t)$ relative to clocks with small $a_x$ and large $a_y$ (\textit{top}). Consequently, we use the population variance along the direction of motion as an instantaneous measure of time-telling ability in the paper.}
\label{fig:MIMotion}
\end{center}
\end{figure}

To see this, we write mutual information between clock state $\vec{c}$ and time $t$ as,
\begin{equation}
MI(C; T) = H(T) - H(T\vert C).
\end{equation}
Here $H(T)$ is a constant, independent of the clock mechanism. Thus, $MI$ depends entirely on the entropy of the distribution $p(t|c)$ of real times given clock state $c$, averaged over different clock states,
\begin{eqnarray}
H(T \vert C) &=& \int p(c) dc H(T|c) \\
 &=& - \int p(c) dc \left[ \int dt p(t|c) \log p(t|c) \right] 
\end{eqnarray}

Consider a clock whose state-space is two dimensional with a periodic x-axis as shown in Fig.\ref{fig:MIMotion}. Further, assume that the distribution $p(\vec{c} \vert t)$ of clock states at a given time is supported on a rectangle of size $a_x \times a_y$ as shown in Fig.\ref{fig:MIMotion} and that the clock states move along the x-axis at a uniform velocity $u$. This situation implies that
\begin{align*}
p(t|c) = \begin{cases}
0 &\text{for  $|c_x - u t| > a_x$}\\
\frac{u}{2 a_x} &\text{for  $|c_x - u t| \leq a_x$}
\end{cases}
\end{align*}
So, 
\begin{align*}
H(T|C)
&= - \int p(c) dc \int_{t = (c_x - a_x)/u}^{(a_x+c_x)/u} dt \frac{u}{2 a_x} \log \left(\frac{u}{2 a_x}\right)\\
&= \log \left(\frac{2 a_x}{u}\right)
\end{align*}
Since $MI(C; T) = H(T) - H(T \vert C)$, $MI$ depends on $- \log a_x$ and is independent of $a_y$, meaning that only the spread in the direction of motion $a_x$ affect the mutual information. Consequently, to understand the quality of time-telling at different times of the day, we project the population variance of $p(\vec{c}\vert t)$ to the direction of the instantaneous velocity of the center of mass of $p(\vec{c}\vert t)$. We use this population variance measure in Figs.3c, e, and 4c, d, e.

\subsection{Cramer-Rao bounds}

Cramer-Rao (CR) bounds quantify the total available information about phase in a given length of history of the signal. Any estimator working with that length of history must necessarily have higher variance (i.e., lower precision) than the Cramer- lower bound corresponding to that length of history. In the limit of infinitely long histories, the CR bound is simply set by the number of bins in time. In our case, this bound is given by $log_2 50 = 5.64$ bits. As shown in Fig. 4, as $L/R \to 0$, limit cycles process longer and longer histories of the external signal. Consequently, the mutual information for such cycles approaches the CR bound  in the limit $L/R \to 0$ as seen in Fig.3f (assuming no internal noise).

\subsection{Hopf bifurcation}
The normal form of the Hopf bifurcation is given by,
\begin{align}
\dot{r} &= \mu \left(r - \frac{r^3}{\mu} \right)  \\
\dot{\theta} &= \omega 
\end{align}

We find limit cycles for $\mu > 0$ which undergo a bifurcation at $\mu = 0$, resulting in point attractors at $\mu < 0$. The dynamics through this bifurcation are characterized by just one parameter, $\mu$, which sets both the radius of the limit cycle ($R \sim \sqrt{\mu}$) and the relaxation time $\tau_{\text{relax}} \sim 1/\mu$ (i.e., the tightness of the quadratic potential around the continuous attractor.) The bifurcation itself occurs at $\mu = 0$.  Consequently, there is no way to interpolate between limit cycles $\mu > 0 $ and point attractors $\mu < 0$ without passing through a region of long relaxation times. Long relaxation times invalidates the models of limit cycles used in this paper; under day-night cycling, limit cycles with long relaxation times lead to orbits that do not visit the attractor at all. That is, the system does not have enough time to relax from the day attractor to the night attractor before the night is over. Consequently, we find that the stable trajectory under cycling conditions is a large orbit that encloses both limit cycles. In such a limit, the continuous attractor of the limit cycle plays no role at all and the limits cycles resemble point attractors. 

Since we seek to contrast the effect of noise on continuous and point attractors (and not the effect of relaxation times), we keep the relaxation time constant in our interpolation. Thus, we use the parametrization,
\begin{align}
\dot{r} &= \alpha \left(r - \frac{r^3}{\mu}\right)  \\
\dot{\theta} &= \omega
\end{align}
where we have two distinct parameters controlling the radius $R \sim \sqrt{\mu}$ and relaxation time $\tau_{relax} \sim \frac{1}{\alpha}$, the latter of which is held constant. This parameterization does have the downside of being singular when $R \sim \sqrt{\mu} \to 0$. Hence we use this parameterization and stay in the regime $R/L > 0.5$ to avoid the singularity at $R=0$.  As seen in Fig.3f, 4f, interpolating down to $R/L \sim 0.5$ already reveals point attractor-like behavior.

\bibliography{arXiv_PRCs_Optimal_Estimation}

\begin{thebibliography}{54}%
\makeatletter
\providecommand \@ifxundefined [1]{%
 \@ifx{#1\undefined}
}%
\providecommand \@ifnum [1]{%
 \ifnum #1\expandafter \@firstoftwo
 \else \expandafter \@secondoftwo
 \fi
}%
\providecommand \@ifx [1]{%
 \ifx #1\expandafter \@firstoftwo
 \else \expandafter \@secondoftwo
 \fi
}%
\providecommand \natexlab [1]{#1}%
\providecommand \enquote  [1]{``#1''}%
\providecommand \bibnamefont  [1]{#1}%
\providecommand \bibfnamefont [1]{#1}%
\providecommand \citenamefont [1]{#1}%
\providecommand \href@noop [0]{\@secondoftwo}%
\providecommand \href [0]{\begingroup \@sanitize@url \@href}%
\providecommand \@href[1]{\@@startlink{#1}\@@href}%
\providecommand \@@href[1]{\endgroup#1\@@endlink}%
\providecommand \@sanitize@url [0]{\catcode `\\12\catcode `\$12\catcode
  `\&12\catcode `\#12\catcode `\^12\catcode `\_12\catcode `\%12\relax}%
\providecommand \@@startlink[1]{}%
\providecommand \@@endlink[0]{}%
\providecommand \url  [0]{\begingroup\@sanitize@url \@url }%
\providecommand \@url [1]{\endgroup\@href {#1}{\urlprefix }}%
\providecommand \urlprefix  [0]{URL }%
\providecommand \Eprint [0]{\href }%
\providecommand \doibase [0]{http://dx.doi.org/}%
\providecommand \selectlanguage [0]{\@gobble}%
\providecommand \bibinfo  [0]{\@secondoftwo}%
\providecommand \bibfield  [0]{\@secondoftwo}%
\providecommand \translation [1]{[#1]}%
\providecommand \BibitemOpen [0]{}%
\providecommand \bibitemStop [0]{}%
\providecommand \bibitemNoStop [0]{.\EOS\space}%
\providecommand \EOS [0]{\spacefactor3000\relax}%
\providecommand \BibitemShut  [1]{\csname bibitem#1\endcsname}%
\let\auto@bib@innerbib\@empty
%</preamble>
\bibitem [{\citenamefont {Bowsher}\ and\ \citenamefont
  {Swain}(2014)}]{Bowsher2014-xx}%
  \BibitemOpen
  \bibfield  {author} {\bibinfo {author} {\bibfnamefont {C.~G.}\ \bibnamefont
  {Bowsher}}\ and\ \bibinfo {author} {\bibfnamefont {P.~S.}\ \bibnamefont
  {Swain}},\ }\href@noop {} {\bibfield  {journal} {\bibinfo  {journal} {Curr.
  Opin. Biotechnol.}\ }\textbf {\bibinfo {volume} {28}},\ \bibinfo {pages}
  {149} (\bibinfo {year} {2014})}\BibitemShut {NoStop}%
\bibitem [{\citenamefont {Mitchell}\ \emph {et~al.}(2015)\citenamefont
  {Mitchell}, \citenamefont {Wei},\ and\ \citenamefont
  {Lim}}]{Mitchell2015-oa}%
  \BibitemOpen
  \bibfield  {author} {\bibinfo {author} {\bibfnamefont {A.}~\bibnamefont
  {Mitchell}}, \bibinfo {author} {\bibfnamefont {P.}~\bibnamefont {Wei}}, \
  and\ \bibinfo {author} {\bibfnamefont {W.~A.}\ \bibnamefont {Lim}},\
  }\href@noop {} {\bibfield  {journal} {\bibinfo  {journal} {Science}\ }\textbf
  {\bibinfo {volume} {350}},\ \bibinfo {pages} {1379} (\bibinfo {year}
  {2015})}\BibitemShut {NoStop}%
\bibitem [{\citenamefont {Sourjik}\ and\ \citenamefont
  {Wingreen}(2012)}]{Sourjik2012-fc}%
  \BibitemOpen
  \bibfield  {author} {\bibinfo {author} {\bibfnamefont {V.}~\bibnamefont
  {Sourjik}}\ and\ \bibinfo {author} {\bibfnamefont {N.~S.}\ \bibnamefont
  {Wingreen}},\ }\href@noop {} {\bibfield  {journal} {\bibinfo  {journal}
  {Curr. Opin. Cell Biol.}\ }\textbf {\bibinfo {volume} {24}},\ \bibinfo
  {pages} {262} (\bibinfo {year} {2012})}\BibitemShut {NoStop}%
\bibitem [{\citenamefont {Tu}\ \emph {et~al.}(2008)\citenamefont {Tu},
  \citenamefont {Shimizu},\ and\ \citenamefont {Berg}}]{Tu2008-dm}%
  \BibitemOpen
  \bibfield  {author} {\bibinfo {author} {\bibfnamefont {Y.}~\bibnamefont
  {Tu}}, \bibinfo {author} {\bibfnamefont {T.~S.}\ \bibnamefont {Shimizu}}, \
  and\ \bibinfo {author} {\bibfnamefont {H.~C.}\ \bibnamefont {Berg}},\
  }\href@noop {} {\bibfield  {journal} {\bibinfo  {journal} {Proc. Natl. Acad.
  Sci. U. S. A.}\ }\textbf {\bibinfo {volume} {105}},\ \bibinfo {pages} {14855}
  (\bibinfo {year} {2008})}\BibitemShut {NoStop}%
\bibitem [{\citenamefont {Cai}\ \emph {et~al.}(2014)\citenamefont {Cai},
  \citenamefont {Katoh-Kurasawa}, \citenamefont {Muramoto}, \citenamefont
  {Santhanam}, \citenamefont {Long}, \citenamefont {Li}, \citenamefont {Ueda},
  \citenamefont {Iglesias}, \citenamefont {Shaulsky},\ and\ \citenamefont
  {Devreotes}}]{Cai2014-ca}%
  \BibitemOpen
  \bibfield  {author} {\bibinfo {author} {\bibfnamefont {H.}~\bibnamefont
  {Cai}}, \bibinfo {author} {\bibfnamefont {M.}~\bibnamefont {Katoh-Kurasawa}},
  \bibinfo {author} {\bibfnamefont {T.}~\bibnamefont {Muramoto}}, \bibinfo
  {author} {\bibfnamefont {B.}~\bibnamefont {Santhanam}}, \bibinfo {author}
  {\bibfnamefont {Y.}~\bibnamefont {Long}}, \bibinfo {author} {\bibfnamefont
  {L.}~\bibnamefont {Li}}, \bibinfo {author} {\bibfnamefont {M.}~\bibnamefont
  {Ueda}}, \bibinfo {author} {\bibfnamefont {P.~A.}\ \bibnamefont {Iglesias}},
  \bibinfo {author} {\bibfnamefont {G.}~\bibnamefont {Shaulsky}}, \ and\
  \bibinfo {author} {\bibfnamefont {P.~N.}\ \bibnamefont {Devreotes}},\
  }\href@noop {} {\bibfield  {journal} {\bibinfo  {journal} {Science}\ }\textbf
  {\bibinfo {volume} {343}},\ \bibinfo {pages} {1249531} (\bibinfo {year}
  {2014})}\BibitemShut {NoStop}%
\bibitem [{\citenamefont {Siggia}\ and\ \citenamefont
  {Vergassola}(2013)}]{Siggia2013-la}%
  \BibitemOpen
  \bibfield  {author} {\bibinfo {author} {\bibfnamefont {E.~D.}\ \bibnamefont
  {Siggia}}\ and\ \bibinfo {author} {\bibfnamefont {M.}~\bibnamefont
  {Vergassola}},\ }\href@noop {} {\bibfield  {journal} {\bibinfo  {journal}
  {Proceedings of the National Academy of Sciences}\ }\textbf {\bibinfo
  {volume} {110}},\ \bibinfo {pages} {E3704} (\bibinfo {year}
  {2013})}\BibitemShut {NoStop}%
\bibitem [{\citenamefont {Mora}\ and\ \citenamefont
  {Wingreen}(2010)}]{Mora2010-tu}%
  \BibitemOpen
  \bibfield  {author} {\bibinfo {author} {\bibfnamefont {T.}~\bibnamefont
  {Mora}}\ and\ \bibinfo {author} {\bibfnamefont {N.~S.}\ \bibnamefont
  {Wingreen}},\ }\href@noop {} {\bibfield  {journal} {\bibinfo  {journal}
  {Phys. Rev. Lett.}\ }\textbf {\bibinfo {volume} {104}},\ \bibinfo {pages}
  {248101} (\bibinfo {year} {2010})}\BibitemShut {NoStop}%
\bibitem [{\citenamefont {Endres}\ and\ \citenamefont
  {Wingreen}(2009)}]{Endres2009-ft}%
  \BibitemOpen
  \bibfield  {author} {\bibinfo {author} {\bibfnamefont {R.~G.}\ \bibnamefont
  {Endres}}\ and\ \bibinfo {author} {\bibfnamefont {N.~S.}\ \bibnamefont
  {Wingreen}},\ }\href@noop {} {\bibfield  {journal} {\bibinfo  {journal}
  {Phys. Rev. Lett.}\ }\textbf {\bibinfo {volume} {103}},\ \bibinfo {pages}
  {158101} (\bibinfo {year} {2009})}\BibitemShut {NoStop}%
\bibitem [{\citenamefont {Winfree}(2001)}]{Winfree2001-pr}%
  \BibitemOpen
  \bibfield  {author} {\bibinfo {author} {\bibfnamefont {A.~T.}\ \bibnamefont
  {Winfree}},\ }\href@noop {} {\emph {\bibinfo {title} {The Geometry of
  Biological Time}}}\ (\bibinfo  {publisher} {Springer Science \& Business
  Media},\ \bibinfo {year} {2001})\BibitemShut {NoStop}%
\bibitem [{\citenamefont {Woelfle}\ \emph {et~al.}(2004)\citenamefont
  {Woelfle}, \citenamefont {Ouyang}, \citenamefont {Phanvijhitsiri},\ and\
  \citenamefont {Johnson}}]{Woelfle2004-bc}%
  \BibitemOpen
  \bibfield  {author} {\bibinfo {author} {\bibfnamefont {M.~A.}\ \bibnamefont
  {Woelfle}}, \bibinfo {author} {\bibfnamefont {Y.}~\bibnamefont {Ouyang}},
  \bibinfo {author} {\bibfnamefont {K.}~\bibnamefont {Phanvijhitsiri}}, \ and\
  \bibinfo {author} {\bibfnamefont {C.~H.}\ \bibnamefont {Johnson}},\
  }\href@noop {} {\bibfield  {journal} {\bibinfo  {journal} {Curr. Biol.}\
  }\textbf {\bibinfo {volume} {14}},\ \bibinfo {pages} {1481} (\bibinfo {year}
  {2004})}\BibitemShut {NoStop}%
\bibitem [{\citenamefont {Quinn}\ and\ \citenamefont
  {Hannan}(2001)}]{Quinn2001-um}%
  \BibitemOpen
  \bibfield  {author} {\bibinfo {author} {\bibfnamefont {B.~G.}\ \bibnamefont
  {Quinn}}\ and\ \bibinfo {author} {\bibfnamefont {E.~J.}\ \bibnamefont
  {Hannan}},\ }\href@noop {} {\emph {\bibinfo {title} {The Estimation and
  Tracking of Frequency}}}\ (\bibinfo  {publisher} {Cambridge University
  Press},\ \bibinfo {year} {2001})\BibitemShut {NoStop}%
\bibitem [{\citenamefont {Liao}(2011)}]{Liao2011-ea}%
  \BibitemOpen
  \bibfield  {author} {\bibinfo {author} {\bibfnamefont {Y.}~\bibnamefont
  {Liao}},\ }\emph {\bibinfo {title} {Phase and Frequency Estimation:
  {High-Accuracy} and {Low-Complexity} Techniques}},\ \href@noop {} {Ph.D.
  thesis},\ \bibinfo  {school} {Worcester Polytechnic Institute} (\bibinfo
  {year} {2011})\BibitemShut {NoStop}%
\bibitem [{\citenamefont {Lovell}\ and\ \citenamefont
  {Williamson}(1992)}]{Lovell1992-sf}%
  \BibitemOpen
  \bibfield  {author} {\bibinfo {author} {\bibfnamefont {B.~C.}\ \bibnamefont
  {Lovell}}\ and\ \bibinfo {author} {\bibfnamefont {R.~C.}\ \bibnamefont
  {Williamson}},\ }\href@noop {} {\bibfield  {journal} {\bibinfo  {journal}
  {IEEE Trans. Signal Process.}\ }\textbf {\bibinfo {volume} {40}},\ \bibinfo
  {pages} {1708} (\bibinfo {year} {1992})}\BibitemShut {NoStop}%
\bibitem [{\citenamefont {Kay}(1989)}]{Kay1989-ua}%
  \BibitemOpen
  \bibfield  {author} {\bibinfo {author} {\bibfnamefont {S.}~\bibnamefont
  {Kay}},\ }\href@noop {} {\bibfield  {journal} {\bibinfo  {journal} {IEEE
  Trans. Acoust.}\ }\textbf {\bibinfo {volume} {37}},\ \bibinfo {pages} {1987}
  (\bibinfo {year} {1989})}\BibitemShut {NoStop}%
\bibitem [{\citenamefont {Tretter}(1985)}]{Tretter1985-bw}%
  \BibitemOpen
  \bibfield  {author} {\bibinfo {author} {\bibfnamefont {S.}~\bibnamefont
  {Tretter}},\ }\href@noop {} {\bibfield  {journal} {\bibinfo  {journal} {IEEE
  Trans. Inf. Theory}\ }\textbf {\bibinfo {volume} {31}},\ \bibinfo {pages}
  {832} (\bibinfo {year} {1985})}\BibitemShut {NoStop}%
\bibitem [{\citenamefont {Bryant}(2003)}]{Bryant2003-zw}%
  \BibitemOpen
  \bibfield  {author} {\bibinfo {author} {\bibfnamefont {D.~A.}\ \bibnamefont
  {Bryant}},\ }\href@noop {} {\bibfield  {journal} {\bibinfo  {journal} {Proc.
  Natl. Acad. Sci. U. S. A.}\ }\textbf {\bibinfo {volume} {100}},\ \bibinfo
  {pages} {9647} (\bibinfo {year} {2003})}\BibitemShut {NoStop}%
\bibitem [{\citenamefont {Gutu}\ and\ \citenamefont
  {O'Shea}(2013)}]{Gutu2013-oy}%
  \BibitemOpen
  \bibfield  {author} {\bibinfo {author} {\bibfnamefont {A.}~\bibnamefont
  {Gutu}}\ and\ \bibinfo {author} {\bibfnamefont {E.~K.}\ \bibnamefont
  {O'Shea}},\ }\href@noop {} {\bibfield  {journal} {\bibinfo  {journal} {Mol.
  Cell}\ }\textbf {\bibinfo {volume} {50}},\ \bibinfo {pages} {288} (\bibinfo
  {year} {2013})}\BibitemShut {NoStop}%
\bibitem [{\citenamefont {Holtzendorff}\ \emph {et~al.}(2008)\citenamefont
  {Holtzendorff}, \citenamefont {Partensky}, \citenamefont {Mella},
  \citenamefont {Lennon}, \citenamefont {Hess},\ and\ \citenamefont
  {Garczarek}}]{Holtzendorff2008-bj}%
  \BibitemOpen
  \bibfield  {author} {\bibinfo {author} {\bibfnamefont {J.}~\bibnamefont
  {Holtzendorff}}, \bibinfo {author} {\bibfnamefont {F.}~\bibnamefont
  {Partensky}}, \bibinfo {author} {\bibfnamefont {D.}~\bibnamefont {Mella}},
  \bibinfo {author} {\bibfnamefont {J.-F.}\ \bibnamefont {Lennon}}, \bibinfo
  {author} {\bibfnamefont {W.~R.}\ \bibnamefont {Hess}}, \ and\ \bibinfo
  {author} {\bibfnamefont {L.}~\bibnamefont {Garczarek}},\ }\href@noop {}
  {\bibfield  {journal} {\bibinfo  {journal} {J. Biol. Rhythms}\ }\textbf
  {\bibinfo {volume} {23}},\ \bibinfo {pages} {187} (\bibinfo {year}
  {2008})}\BibitemShut {NoStop}%
\bibitem [{\citenamefont {Dufresne}\ \emph {et~al.}(2003)\citenamefont
  {Dufresne}, \citenamefont {Salanoubat}, \citenamefont {Partensky},
  \citenamefont {Artiguenave}, \citenamefont {Axmann}, \citenamefont {Barbe},
  \citenamefont {Duprat}, \citenamefont {Galperin}, \citenamefont {Koonin},
  \citenamefont {Le~Gall}, \citenamefont {Makarova}, \citenamefont {Ostrowski},
  \citenamefont {Oztas}, \citenamefont {Robert}, \citenamefont {Rogozin},
  \citenamefont {Scanlan}, \citenamefont {Tandeau~de Marsac}, \citenamefont
  {Weissenbach}, \citenamefont {Wincker}, \citenamefont {Wolf},\ and\
  \citenamefont {Hess}}]{Dufresne2003-gh}%
  \BibitemOpen
  \bibfield  {author} {\bibinfo {author} {\bibfnamefont {A.}~\bibnamefont
  {Dufresne}}, \bibinfo {author} {\bibfnamefont {M.}~\bibnamefont
  {Salanoubat}}, \bibinfo {author} {\bibfnamefont {F.}~\bibnamefont
  {Partensky}}, \bibinfo {author} {\bibfnamefont {F.}~\bibnamefont
  {Artiguenave}}, \bibinfo {author} {\bibfnamefont {I.~M.}\ \bibnamefont
  {Axmann}}, \bibinfo {author} {\bibfnamefont {V.}~\bibnamefont {Barbe}},
  \bibinfo {author} {\bibfnamefont {S.}~\bibnamefont {Duprat}}, \bibinfo
  {author} {\bibfnamefont {M.~Y.}\ \bibnamefont {Galperin}}, \bibinfo {author}
  {\bibfnamefont {E.~V.}\ \bibnamefont {Koonin}}, \bibinfo {author}
  {\bibfnamefont {F.}~\bibnamefont {Le~Gall}}, \bibinfo {author} {\bibfnamefont
  {K.~S.}\ \bibnamefont {Makarova}}, \bibinfo {author} {\bibfnamefont
  {M.}~\bibnamefont {Ostrowski}}, \bibinfo {author} {\bibfnamefont
  {S.}~\bibnamefont {Oztas}}, \bibinfo {author} {\bibfnamefont
  {C.}~\bibnamefont {Robert}}, \bibinfo {author} {\bibfnamefont {I.~B.}\
  \bibnamefont {Rogozin}}, \bibinfo {author} {\bibfnamefont {D.~J.}\
  \bibnamefont {Scanlan}}, \bibinfo {author} {\bibfnamefont {N.}~\bibnamefont
  {Tandeau~de Marsac}}, \bibinfo {author} {\bibfnamefont {J.}~\bibnamefont
  {Weissenbach}}, \bibinfo {author} {\bibfnamefont {P.}~\bibnamefont
  {Wincker}}, \bibinfo {author} {\bibfnamefont {Y.~I.}\ \bibnamefont {Wolf}}, \
  and\ \bibinfo {author} {\bibfnamefont {W.~R.}\ \bibnamefont {Hess}},\
  }\href@noop {} {\bibfield  {journal} {\bibinfo  {journal} {Proc. Natl. Acad.
  Sci. U. S. A.}\ }\textbf {\bibinfo {volume} {100}},\ \bibinfo {pages} {10020}
  (\bibinfo {year} {2003})}\BibitemShut {NoStop}%
\bibitem [{\citenamefont {Kitayama}\ \emph {et~al.}(2003)\citenamefont
  {Kitayama}, \citenamefont {Iwasaki}, \citenamefont {Nishiwaki},\ and\
  \citenamefont {Kondo}}]{Kitayama2003-na}%
  \BibitemOpen
  \bibfield  {author} {\bibinfo {author} {\bibfnamefont {Y.}~\bibnamefont
  {Kitayama}}, \bibinfo {author} {\bibfnamefont {H.}~\bibnamefont {Iwasaki}},
  \bibinfo {author} {\bibfnamefont {T.}~\bibnamefont {Nishiwaki}}, \ and\
  \bibinfo {author} {\bibfnamefont {T.}~\bibnamefont {Kondo}},\ }\href@noop {}
  {\bibfield  {journal} {\bibinfo  {journal} {EMBO J.}\ }\textbf {\bibinfo
  {volume} {22}},\ \bibinfo {pages} {2127} (\bibinfo {year}
  {2003})}\BibitemShut {NoStop}%
\bibitem [{\citenamefont {Potoyan}\ and\ \citenamefont
  {Wolynes}(2014)}]{Potoyan2014-so}%
  \BibitemOpen
  \bibfield  {author} {\bibinfo {author} {\bibfnamefont {D.~A.}\ \bibnamefont
  {Potoyan}}\ and\ \bibinfo {author} {\bibfnamefont {P.~G.}\ \bibnamefont
  {Wolynes}},\ }\href@noop {} {\bibfield  {journal} {\bibinfo  {journal}
  {Proceedings of the National Academy of Sciences}\ }\textbf {\bibinfo
  {volume} {111}},\ \bibinfo {pages} {2391} (\bibinfo {year}
  {2014})}\BibitemShut {NoStop}%
\bibitem [{\citenamefont {Laughlin}(1981)}]{Laughlin1981-hz}%
  \BibitemOpen
  \bibfield  {author} {\bibinfo {author} {\bibfnamefont {S.}~\bibnamefont
  {Laughlin}},\ }\href@noop {} {\bibfield  {journal} {\bibinfo  {journal} {Z.
  Naturforsch. C}\ }\textbf {\bibinfo {volume} {36}},\ \bibinfo {pages} {910}
  (\bibinfo {year} {1981})}\BibitemShut {NoStop}%
\bibitem [{\citenamefont {Fu}\ and\ \citenamefont {Kam}(2007)}]{Fu2007-ql}%
  \BibitemOpen
  \bibfield  {author} {\bibinfo {author} {\bibfnamefont {H.}~\bibnamefont
  {Fu}}\ and\ \bibinfo {author} {\bibfnamefont {P.~Y.}\ \bibnamefont {Kam}},\
  }\href@noop {} {\bibfield  {journal} {\bibinfo  {journal} {IEEE Trans. Signal
  Process.}\ }\textbf {\bibinfo {volume} {55}},\ \bibinfo {pages} {834}
  (\bibinfo {year} {2007})}\BibitemShut {NoStop}%
\bibitem [{\citenamefont {Ghogho}\ \emph {et~al.}(1999)\citenamefont {Ghogho},
  \citenamefont {{Member}}, \citenamefont {Nandi}, \citenamefont {{Senior
  Member}}, \citenamefont {Swami},\ and\ \citenamefont {{Senior
  Member}}}]{Ghogho1999-ni}%
  \BibitemOpen
  \bibfield  {author} {\bibinfo {author} {\bibfnamefont {M.}~\bibnamefont
  {Ghogho}}, \bibinfo {author} {\bibnamefont {{Member}}}, \bibinfo {author}
  {\bibfnamefont {A.~K.}\ \bibnamefont {Nandi}}, \bibinfo {author}
  {\bibnamefont {{Senior Member}}}, \bibinfo {author} {\bibfnamefont
  {A.}~\bibnamefont {Swami}}, \ and\ \bibinfo {author} {\bibnamefont {{Senior
  Member}}},\ }\href@noop {} {\bibfield  {journal} {\bibinfo  {journal} {IEEE
  Trans. Signal Process.}\ }\textbf {\bibinfo {volume} {47}} (\bibinfo {year}
  {1999})}\BibitemShut {NoStop}%
\bibitem [{\citenamefont {Rust}\ \emph {et~al.}(2007)\citenamefont {Rust},
  \citenamefont {Markson}, \citenamefont {Lane}, \citenamefont {Fisher},\ and\
  \citenamefont {O'Shea}}]{Rust2007-op}%
  \BibitemOpen
  \bibfield  {author} {\bibinfo {author} {\bibfnamefont {M.~J.}\ \bibnamefont
  {Rust}}, \bibinfo {author} {\bibfnamefont {J.~S.}\ \bibnamefont {Markson}},
  \bibinfo {author} {\bibfnamefont {W.~S.}\ \bibnamefont {Lane}}, \bibinfo
  {author} {\bibfnamefont {D.~S.}\ \bibnamefont {Fisher}}, \ and\ \bibinfo
  {author} {\bibfnamefont {E.~K.}\ \bibnamefont {O'Shea}},\ }\href@noop {}
  {\bibfield  {journal} {\bibinfo  {journal} {Science}\ }\textbf {\bibinfo
  {volume} {318}},\ \bibinfo {pages} {809} (\bibinfo {year}
  {2007})}\BibitemShut {NoStop}%
\bibitem [{\citenamefont {Leypunskiy}\ \emph {et~al.}(2017)\citenamefont
  {Leypunskiy}, \citenamefont {Lin}, \citenamefont {Yoo}, \citenamefont {Lee},
  \citenamefont {Dinner},\ and\ \citenamefont {Rust}}]{Leypunskiy2017-al}%
  \BibitemOpen
  \bibfield  {author} {\bibinfo {author} {\bibfnamefont {E.}~\bibnamefont
  {Leypunskiy}}, \bibinfo {author} {\bibfnamefont {J.}~\bibnamefont {Lin}},
  \bibinfo {author} {\bibfnamefont {H.}~\bibnamefont {Yoo}}, \bibinfo {author}
  {\bibfnamefont {U.}~\bibnamefont {Lee}}, \bibinfo {author} {\bibfnamefont
  {A.~R.}\ \bibnamefont {Dinner}}, \ and\ \bibinfo {author} {\bibfnamefont
  {M.~J.}\ \bibnamefont {Rust}},\ }\href@noop {} {\bibfield  {journal}
  {\bibinfo  {journal} {Elife}\ }\textbf {\bibinfo {volume} {6}} (\bibinfo
  {year} {2017})}\BibitemShut {NoStop}%
\bibitem [{\citenamefont {Pattanayak}\ \emph {et~al.}(2014)\citenamefont
  {Pattanayak}, \citenamefont {Phong},\ and\ \citenamefont
  {Rust}}]{Pattanayak2014-bv}%
  \BibitemOpen
  \bibfield  {author} {\bibinfo {author} {\bibfnamefont {G.~K.}\ \bibnamefont
  {Pattanayak}}, \bibinfo {author} {\bibfnamefont {C.}~\bibnamefont {Phong}}, \
  and\ \bibinfo {author} {\bibfnamefont {M.~J.}\ \bibnamefont {Rust}},\
  }\href@noop {} {\bibfield  {journal} {\bibinfo  {journal} {Curr. Biol.}\
  }\textbf {\bibinfo {volume} {24}},\ \bibinfo {pages} {1934} (\bibinfo {year}
  {2014})}\BibitemShut {NoStop}%
\bibitem [{\citenamefont {Zwicker}\ \emph {et~al.}(2010)\citenamefont
  {Zwicker}, \citenamefont {Lubensky},\ and\ \citenamefont {ten
  Wolde}}]{Zwicker2010-de}%
  \BibitemOpen
  \bibfield  {author} {\bibinfo {author} {\bibfnamefont {D.}~\bibnamefont
  {Zwicker}}, \bibinfo {author} {\bibfnamefont {D.~K.}\ \bibnamefont
  {Lubensky}}, \ and\ \bibinfo {author} {\bibfnamefont {P.~R.}\ \bibnamefont
  {ten Wolde}},\ }\href@noop {} {\bibfield  {journal} {\bibinfo  {journal}
  {Proceedings of the National Academy of Sciences}\ }\textbf {\bibinfo
  {volume} {107}},\ \bibinfo {pages} {22540} (\bibinfo {year}
  {2010})}\BibitemShut {NoStop}%
\bibitem [{\citenamefont {Paijmans}\ \emph {et~al.}(2016)\citenamefont
  {Paijmans}, \citenamefont {Bosman}, \citenamefont {ten Wolde},\ and\
  \citenamefont {Lubensky}}]{Paijmans2016-hs}%
  \BibitemOpen
  \bibfield  {author} {\bibinfo {author} {\bibfnamefont {J.}~\bibnamefont
  {Paijmans}}, \bibinfo {author} {\bibfnamefont {M.}~\bibnamefont {Bosman}},
  \bibinfo {author} {\bibfnamefont {P.~R.}\ \bibnamefont {ten Wolde}}, \ and\
  \bibinfo {author} {\bibfnamefont {D.~K.}\ \bibnamefont {Lubensky}},\
  }\href@noop {} {\bibfield  {journal} {\bibinfo  {journal} {Proceedings of the
  National Academy of Sciences}\ }\textbf {\bibinfo {volume} {113}},\ \bibinfo
  {pages} {4063} (\bibinfo {year} {2016})}\BibitemShut {NoStop}%
\bibitem [{\citenamefont {Heltberg}\ \emph {et~al.}(2016)\citenamefont
  {Heltberg}, \citenamefont {Kellogg}, \citenamefont {Krishna}, \citenamefont
  {Tay},\ and\ \citenamefont {Jensen}}]{Heltberg2016-ot}%
  \BibitemOpen
  \bibfield  {author} {\bibinfo {author} {\bibfnamefont {M.}~\bibnamefont
  {Heltberg}}, \bibinfo {author} {\bibfnamefont {R.~A.}\ \bibnamefont
  {Kellogg}}, \bibinfo {author} {\bibfnamefont {S.}~\bibnamefont {Krishna}},
  \bibinfo {author} {\bibfnamefont {S.}~\bibnamefont {Tay}}, \ and\ \bibinfo
  {author} {\bibfnamefont {M.~H.}\ \bibnamefont {Jensen}},\ }\href@noop {}
  {\bibfield  {journal} {\bibinfo  {journal} {Cell Syst}\ }\textbf {\bibinfo
  {volume} {3}},\ \bibinfo {pages} {532} (\bibinfo {year} {2016})}\BibitemShut
  {NoStop}%
\bibitem [{\citenamefont {Potvin-Trottier}\ \emph {et~al.}(2016)\citenamefont
  {Potvin-Trottier}, \citenamefont {Lord}, \citenamefont {Vinnicombe},\ and\
  \citenamefont {Paulsson}}]{Potvin-Trottier2016-bz}%
  \BibitemOpen
  \bibfield  {author} {\bibinfo {author} {\bibfnamefont {L.}~\bibnamefont
  {Potvin-Trottier}}, \bibinfo {author} {\bibfnamefont {N.~D.}\ \bibnamefont
  {Lord}}, \bibinfo {author} {\bibfnamefont {G.}~\bibnamefont {Vinnicombe}}, \
  and\ \bibinfo {author} {\bibfnamefont {J.}~\bibnamefont {Paulsson}},\
  }\href@noop {} {\bibfield  {journal} {\bibinfo  {journal} {Nature}\ }\textbf
  {\bibinfo {volume} {538}},\ \bibinfo {pages} {514} (\bibinfo {year}
  {2016})}\BibitemShut {NoStop}%
\bibitem [{\citenamefont {Tsai}\ \emph {et~al.}(2008)\citenamefont {Tsai},
  \citenamefont {Choi}, \citenamefont {Ma}, \citenamefont {Pomerening},
  \citenamefont {Tang},\ and\ \citenamefont {Ferrell}}]{Tsai2008-gg}%
  \BibitemOpen
  \bibfield  {author} {\bibinfo {author} {\bibfnamefont {T.~Y.-C.}\
  \bibnamefont {Tsai}}, \bibinfo {author} {\bibfnamefont {Y.~S.}\ \bibnamefont
  {Choi}}, \bibinfo {author} {\bibfnamefont {W.}~\bibnamefont {Ma}}, \bibinfo
  {author} {\bibfnamefont {J.~R.}\ \bibnamefont {Pomerening}}, \bibinfo
  {author} {\bibfnamefont {C.}~\bibnamefont {Tang}}, \ and\ \bibinfo {author}
  {\bibfnamefont {J.~E.}\ \bibnamefont {Ferrell}, \bibfnamefont {Jr}},\
  }\href@noop {} {\bibfield  {journal} {\bibinfo  {journal} {Science}\ }\textbf
  {\bibinfo {volume} {321}},\ \bibinfo {pages} {126} (\bibinfo {year}
  {2008})}\BibitemShut {NoStop}%
\bibitem [{\citenamefont {Elowitz}\ and\ \citenamefont
  {Leibler}(2000)}]{Elowitz2000-lz}%
  \BibitemOpen
  \bibfield  {author} {\bibinfo {author} {\bibfnamefont {M.~B.}\ \bibnamefont
  {Elowitz}}\ and\ \bibinfo {author} {\bibfnamefont {S.}~\bibnamefont
  {Leibler}},\ }\href@noop {} {\bibfield  {journal} {\bibinfo  {journal}
  {Nature}\ }\textbf {\bibinfo {volume} {403}},\ \bibinfo {pages} {335}
  (\bibinfo {year} {2000})}\BibitemShut {NoStop}%
\bibitem [{\citenamefont {Saunders}(2002)}]{Saunders2002-hj}%
  \BibitemOpen
  \bibfield  {author} {\bibinfo {author} {\bibfnamefont {D.~S.}\ \bibnamefont
  {Saunders}},\ }\href@noop {} {\emph {\bibinfo {title} {Insect Clocks, Third
  Edition}}}\ (\bibinfo  {publisher} {Elsevier},\ \bibinfo {year}
  {2002})\BibitemShut {NoStop}%
\bibitem [{\citenamefont {Murayama}\ \emph {et~al.}(2017)\citenamefont
  {Murayama}, \citenamefont {Kori}, \citenamefont {Oshima}, \citenamefont
  {Kondo}, \citenamefont {Iwasaki},\ and\ \citenamefont
  {Ito}}]{Murayama2017-vj}%
  \BibitemOpen
  \bibfield  {author} {\bibinfo {author} {\bibfnamefont {Y.}~\bibnamefont
  {Murayama}}, \bibinfo {author} {\bibfnamefont {H.}~\bibnamefont {Kori}},
  \bibinfo {author} {\bibfnamefont {C.}~\bibnamefont {Oshima}}, \bibinfo
  {author} {\bibfnamefont {T.}~\bibnamefont {Kondo}}, \bibinfo {author}
  {\bibfnamefont {H.}~\bibnamefont {Iwasaki}}, \ and\ \bibinfo {author}
  {\bibfnamefont {H.}~\bibnamefont {Ito}},\ }\href@noop {} {\bibfield
  {journal} {\bibinfo  {journal} {Proc. Natl. Acad. Sci. U. S. A.}\ } (\bibinfo
  {year} {2017})}\BibitemShut {NoStop}%
\bibitem [{\citenamefont {Gu}\ \emph {et~al.}(2001)\citenamefont {Gu},
  \citenamefont {Fuentes}, \citenamefont {Garstang}, \citenamefont {Silva},
  \citenamefont {Heitz}, \citenamefont {Sigler},\ and\ \citenamefont
  {Shugart}}]{Gu2001-re}%
  \BibitemOpen
  \bibfield  {author} {\bibinfo {author} {\bibfnamefont {L.}~\bibnamefont
  {Gu}}, \bibinfo {author} {\bibfnamefont {J.~D.}\ \bibnamefont {Fuentes}},
  \bibinfo {author} {\bibfnamefont {M.}~\bibnamefont {Garstang}}, \bibinfo
  {author} {\bibfnamefont {J.~T.~d.}\ \bibnamefont {Silva}}, \bibinfo {author}
  {\bibfnamefont {R.}~\bibnamefont {Heitz}}, \bibinfo {author} {\bibfnamefont
  {J.}~\bibnamefont {Sigler}}, \ and\ \bibinfo {author} {\bibfnamefont {H.~H.}\
  \bibnamefont {Shugart}},\ }\href@noop {} {\bibfield  {journal} {\bibinfo
  {journal} {Agric. For. Meteorol.}\ }\textbf {\bibinfo {volume} {106}},\
  \bibinfo {pages} {117} (\bibinfo {year} {2001})}\BibitemShut {NoStop}%
\bibitem [{\citenamefont {Tostevin}\ and\ \citenamefont {ten
  Wolde}(2009)}]{Tostevin2009-cm}%
  \BibitemOpen
  \bibfield  {author} {\bibinfo {author} {\bibfnamefont {F.}~\bibnamefont
  {Tostevin}}\ and\ \bibinfo {author} {\bibfnamefont {P.~R.}\ \bibnamefont {ten
  Wolde}},\ }\href@noop {} {\bibfield  {journal} {\bibinfo  {journal} {Phys.
  Rev. Lett.}\ }\textbf {\bibinfo {volume} {102}},\ \bibinfo {pages} {218101}
  (\bibinfo {year} {2009})}\BibitemShut {NoStop}%
\bibitem [{\citenamefont {Swain}\ \emph {et~al.}(2002)\citenamefont {Swain},
  \citenamefont {Elowitz},\ and\ \citenamefont {Siggia}}]{Swain2002-tj}%
  \BibitemOpen
  \bibfield  {author} {\bibinfo {author} {\bibfnamefont {P.~S.}\ \bibnamefont
  {Swain}}, \bibinfo {author} {\bibfnamefont {M.~B.}\ \bibnamefont {Elowitz}},
  \ and\ \bibinfo {author} {\bibfnamefont {E.~D.}\ \bibnamefont {Siggia}},\
  }\href@noop {} {\bibfield  {journal} {\bibinfo  {journal} {Proc. Natl. Acad.
  Sci. U. S. A.}\ }\textbf {\bibinfo {volume} {99}},\ \bibinfo {pages} {12795}
  (\bibinfo {year} {2002})}\BibitemShut {NoStop}%
\bibitem [{\citenamefont {Potoyan}\ \emph {et~al.}(2016)\citenamefont
  {Potoyan}, \citenamefont {Zheng}, \citenamefont {Komives},\ and\
  \citenamefont {Wolynes}}]{Potoyan2016-li}%
  \BibitemOpen
  \bibfield  {author} {\bibinfo {author} {\bibfnamefont {D.~A.}\ \bibnamefont
  {Potoyan}}, \bibinfo {author} {\bibfnamefont {W.}~\bibnamefont {Zheng}},
  \bibinfo {author} {\bibfnamefont {E.~A.}\ \bibnamefont {Komives}}, \ and\
  \bibinfo {author} {\bibfnamefont {P.~G.}\ \bibnamefont {Wolynes}},\
  }\href@noop {} {\bibfield  {journal} {\bibinfo  {journal} {Proceedings of the
  National Academy of Sciences}\ }\textbf {\bibinfo {volume} {113}},\ \bibinfo
  {pages} {110} (\bibinfo {year} {2016})}\BibitemShut {NoStop}%
\bibitem [{\citenamefont {Ziv}\ \emph {et~al.}(2007)\citenamefont {Ziv},
  \citenamefont {Nemenman},\ and\ \citenamefont {Wiggins}}]{Ziv2007-ga}%
  \BibitemOpen
  \bibfield  {author} {\bibinfo {author} {\bibfnamefont {E.}~\bibnamefont
  {Ziv}}, \bibinfo {author} {\bibfnamefont {I.}~\bibnamefont {Nemenman}}, \
  and\ \bibinfo {author} {\bibfnamefont {C.~H.}\ \bibnamefont {Wiggins}},\
  }\href@noop {} {\bibfield  {journal} {\bibinfo  {journal} {PLoS One}\
  }\textbf {\bibinfo {volume} {2}},\ \bibinfo {pages} {e1077} (\bibinfo {year}
  {2007})}\BibitemShut {NoStop}%
\bibitem [{\citenamefont {Mugler}\ \emph {et~al.}(2010)\citenamefont {Mugler},
  \citenamefont {Walczak},\ and\ \citenamefont {Wiggins}}]{Mugler2010-bh}%
  \BibitemOpen
  \bibfield  {author} {\bibinfo {author} {\bibfnamefont {A.}~\bibnamefont
  {Mugler}}, \bibinfo {author} {\bibfnamefont {A.~M.}\ \bibnamefont {Walczak}},
  \ and\ \bibinfo {author} {\bibfnamefont {C.~H.}\ \bibnamefont {Wiggins}},\
  }\href@noop {} {\bibfield  {journal} {\bibinfo  {journal} {arXiv.org}\
  }\textbf {\bibinfo {volume} {q-bio.MN}},\ \bibinfo {pages} {058101} (\bibinfo
  {year} {2010})}\BibitemShut {NoStop}%
\bibitem [{\citenamefont {Qian}(2011)}]{Qian2011-yc}%
  \BibitemOpen
  \bibfield  {author} {\bibinfo {author} {\bibfnamefont {H.}~\bibnamefont
  {Qian}},\ }\href@noop {} {\bibfield  {journal} {\bibinfo  {journal}
  {Nonlinearity}\ }\textbf {\bibinfo {volume} {24}},\ \bibinfo {pages} {R19}
  (\bibinfo {year} {2011})}\BibitemShut {NoStop}%
\bibitem [{\citenamefont {Gillespie}(2007)}]{Gillespie2007-yr}%
  \BibitemOpen
  \bibfield  {author} {\bibinfo {author} {\bibfnamefont {D.~T.}\ \bibnamefont
  {Gillespie}},\ }\href@noop {} {\bibfield  {journal} {\bibinfo  {journal}
  {Annu. Rev. Phys. Chem.}\ }\textbf {\bibinfo {volume} {58}},\ \bibinfo
  {pages} {35} (\bibinfo {year} {2007})}\BibitemShut {NoStop}%
\bibitem [{\citenamefont {Mihalcescu}\ \emph {et~al.}(2004)\citenamefont
  {Mihalcescu}, \citenamefont {Hsing},\ and\ \citenamefont
  {Leibler}}]{Mihalcescu2004-ov}%
  \BibitemOpen
  \bibfield  {author} {\bibinfo {author} {\bibfnamefont {I.}~\bibnamefont
  {Mihalcescu}}, \bibinfo {author} {\bibfnamefont {W.}~\bibnamefont {Hsing}}, \
  and\ \bibinfo {author} {\bibfnamefont {S.}~\bibnamefont {Leibler}},\
  }\href@noop {} {\bibfield  {journal} {\bibinfo  {journal} {Nature}\ }\textbf
  {\bibinfo {volume} {430}},\ \bibinfo {pages} {81} (\bibinfo {year}
  {2004})}\BibitemShut {NoStop}%
\bibitem [{\citenamefont {Barkai}\ and\ \citenamefont
  {Leibler}(2000)}]{Barkai2000-nr}%
  \BibitemOpen
  \bibfield  {author} {\bibinfo {author} {\bibfnamefont {N.}~\bibnamefont
  {Barkai}}\ and\ \bibinfo {author} {\bibfnamefont {S.}~\bibnamefont
  {Leibler}},\ }\href@noop {} {\bibfield  {journal} {\bibinfo  {journal}
  {Nature}\ }\textbf {\bibinfo {volume} {403}},\ \bibinfo {pages} {267}
  (\bibinfo {year} {2000})}\BibitemShut {NoStop}%
\bibitem [{\citenamefont {Gonze}\ \emph {et~al.}(2002)\citenamefont {Gonze},
  \citenamefont {Halloy},\ and\ \citenamefont {Goldbeter}}]{Gonze2002-rx}%
  \BibitemOpen
  \bibfield  {author} {\bibinfo {author} {\bibfnamefont {D.}~\bibnamefont
  {Gonze}}, \bibinfo {author} {\bibfnamefont {J.}~\bibnamefont {Halloy}}, \
  and\ \bibinfo {author} {\bibfnamefont {A.}~\bibnamefont {Goldbeter}},\
  }\href@noop {} {\bibfield  {journal} {\bibinfo  {journal} {Proc. Natl. Acad.
  Sci. U. S. A.}\ }\textbf {\bibinfo {volume} {99}},\ \bibinfo {pages} {673}
  (\bibinfo {year} {2002})}\BibitemShut {NoStop}%
\bibitem [{\citenamefont {Cao}\ \emph {et~al.}(2015)\citenamefont {Cao},
  \citenamefont {Wang}, \citenamefont {Ouyang},\ and\ \citenamefont
  {Tu}}]{Cao2015-vz}%
  \BibitemOpen
  \bibfield  {author} {\bibinfo {author} {\bibfnamefont {Y.}~\bibnamefont
  {Cao}}, \bibinfo {author} {\bibfnamefont {H.}~\bibnamefont {Wang}}, \bibinfo
  {author} {\bibfnamefont {Q.}~\bibnamefont {Ouyang}}, \ and\ \bibinfo {author}
  {\bibfnamefont {Y.}~\bibnamefont {Tu}},\ }\href@noop {} {\bibfield  {journal}
  {\bibinfo  {journal} {Nat. Phys.}\ }\textbf {\bibinfo {volume} {11}},\
  \bibinfo {pages} {772} (\bibinfo {year} {2015})}\BibitemShut {NoStop}%
\bibitem [{\citenamefont {Monti}\ \emph {et~al.}(2017)\citenamefont {Monti},
  \citenamefont {Lubensky},\ and\ \citenamefont {ten Wolde}}]{Monti2017-dm}%
  \BibitemOpen
  \bibfield  {author} {\bibinfo {author} {\bibfnamefont {M.}~\bibnamefont
  {Monti}}, \bibinfo {author} {\bibfnamefont {D.~K.}\ \bibnamefont {Lubensky}},
  \ and\ \bibinfo {author} {\bibfnamefont {P.~R.}\ \bibnamefont {ten Wolde}},\
  }\href@noop {} {\  (\bibinfo {year} {2017})}\BibitemShut {NoStop}%
\bibitem [{\citenamefont {Sontag}(2003)}]{Sontag2003-lk}%
  \BibitemOpen
  \bibfield  {author} {\bibinfo {author} {\bibfnamefont {E.~D.}\ \bibnamefont
  {Sontag}},\ }\href@noop {} {\bibfield  {journal} {\bibinfo  {journal} {Syst.
  Control Lett.}\ }\textbf {\bibinfo {volume} {50}},\ \bibinfo {pages} {119}
  (\bibinfo {year} {2003})}\BibitemShut {NoStop}%
\bibitem [{\citenamefont {Burak}\ and\ \citenamefont
  {Fiete}(2012)}]{Burak2012-bu}%
  \BibitemOpen
  \bibfield  {author} {\bibinfo {author} {\bibfnamefont {Y.}~\bibnamefont
  {Burak}}\ and\ \bibinfo {author} {\bibfnamefont {I.~R.}\ \bibnamefont
  {Fiete}},\ }\href@noop {} {\bibfield  {journal} {\bibinfo  {journal} {Proc.
  Natl. Acad. Sci. U. S. A.}\ }\textbf {\bibinfo {volume} {109}},\ \bibinfo
  {pages} {17645} (\bibinfo {year} {2012})}\BibitemShut {NoStop}%
\bibitem [{\citenamefont {Seung}\ \emph {et~al.}(2000)\citenamefont {Seung},
  \citenamefont {Lee}, \citenamefont {Reis},\ and\ \citenamefont
  {Tank}}]{Seung2000-bk}%
  \BibitemOpen
  \bibfield  {author} {\bibinfo {author} {\bibfnamefont {H.~S.}\ \bibnamefont
  {Seung}}, \bibinfo {author} {\bibfnamefont {D.~D.}\ \bibnamefont {Lee}},
  \bibinfo {author} {\bibfnamefont {B.~Y.}\ \bibnamefont {Reis}}, \ and\
  \bibinfo {author} {\bibfnamefont {D.~W.}\ \bibnamefont {Tank}},\ }\href@noop
  {} {\bibfield  {journal} {\bibinfo  {journal} {Neuron}\ }\textbf {\bibinfo
  {volume} {26}},\ \bibinfo {pages} {259} (\bibinfo {year} {2000})}\BibitemShut
  {NoStop}%
\bibitem [{\citenamefont {Yoon}\ \emph {et~al.}(2013)\citenamefont {Yoon},
  \citenamefont {Buice}, \citenamefont {Barry}, \citenamefont {Hayman},
  \citenamefont {Burgess},\ and\ \citenamefont {Fiete}}]{Yoon2013-nl}%
  \BibitemOpen
  \bibfield  {author} {\bibinfo {author} {\bibfnamefont {K.}~\bibnamefont
  {Yoon}}, \bibinfo {author} {\bibfnamefont {M.~A.}\ \bibnamefont {Buice}},
  \bibinfo {author} {\bibfnamefont {C.}~\bibnamefont {Barry}}, \bibinfo
  {author} {\bibfnamefont {R.}~\bibnamefont {Hayman}}, \bibinfo {author}
  {\bibfnamefont {N.}~\bibnamefont {Burgess}}, \ and\ \bibinfo {author}
  {\bibfnamefont {I.~R.}\ \bibnamefont {Fiete}},\ }\href@noop {} {\bibfield
  {journal} {\bibinfo  {journal} {Nat. Neurosci.}\ }\textbf {\bibinfo {volume}
  {16}},\ \bibinfo {pages} {1077} (\bibinfo {year} {2013})}\BibitemShut
  {NoStop}%
\bibitem [{\citenamefont {Vaze}\ and\ \citenamefont
  {Helfrich-F{\"o}rster}(2016)}]{Vaze2016-nj}%
  \BibitemOpen
  \bibfield  {author} {\bibinfo {author} {\bibfnamefont {K.~M.}\ \bibnamefont
  {Vaze}}\ and\ \bibinfo {author} {\bibfnamefont {C.}~\bibnamefont
  {Helfrich-F{\"o}rster}},\ }\href@noop {} {\bibfield  {journal} {\bibinfo
  {journal} {Physiol. Entomol.}\ }\textbf {\bibinfo {volume} {41}},\ \bibinfo
  {pages} {378} (\bibinfo {year} {2016})}\BibitemShut {NoStop}%
\bibitem [{\citenamefont {Kidd}\ \emph {et~al.}(2015)\citenamefont {Kidd},
  \citenamefont {Young},\ and\ \citenamefont {Siggia}}]{Kidd2015-cr}%
  \BibitemOpen
  \bibfield  {author} {\bibinfo {author} {\bibfnamefont {P.~B.}\ \bibnamefont
  {Kidd}}, \bibinfo {author} {\bibfnamefont {M.~W.}\ \bibnamefont {Young}}, \
  and\ \bibinfo {author} {\bibfnamefont {E.~D.}\ \bibnamefont {Siggia}},\
  }\href@noop {} {\bibfield  {journal} {\bibinfo  {journal} {Proc. Natl. Acad.
  Sci. U. S. A.}\ }\textbf {\bibinfo {volume} {112}},\ \bibinfo {pages} {E6284}
  (\bibinfo {year} {2015})}\BibitemShut {NoStop}%
\end{thebibliography}%

\end{document}